\documentclass[aps,twocolumn,showpacs,nofootinbib,aps,longbibliography,superscriptaddress]{revtex4-1}
\usepackage[utf8]{inputenc}
\usepackage{color}
\usepackage{amsmath}
\usepackage{url}
\usepackage{graphicx}
\usepackage{amssymb}
\usepackage[dvipsnames]{xcolor}
\usepackage{appendix}
\usepackage[section]{placeins}
\usepackage{multirow}
\usepackage{hyperref}
\usepackage{mathrsfs} 
\usepackage{bm}
\usepackage{soul}
\usepackage{array}
\usepackage{lipsum}
\usepackage{relsize}
\usepackage{amsfonts}
\usepackage{comment}
\usepackage{outlines}
\usepackage[normalem]{ulem}
\usepackage{multirow}

\usepackage[ruled,lined]{algorithm2e}

\begin{document}

\title{Bayes goes fast: Uncertainty Quantification for a Covariant Energy Density Functional emulated by the Reduced Basis Method} 


\author{Pablo Giuliani}
\altaffiliation{Giuliani and Godbey are co-first authors contributing equally to this work}
\email[\\]{giulianp@frib.msu.edu}
\affiliation{FRIB/NSCL Laboratory, Michigan State University, East Lansing, Michigan 48824, USA}
\affiliation{Department of Statistics and Probability, Michigan State University, East Lansing, Michigan 48824, USA}

\author{Kyle Godbey}
\email{godbey@frib.msu.edu}
\affiliation{FRIB/NSCL Laboratory, Michigan State University, East Lansing, Michigan 48824, USA}

\author{Edgard Bonilla}
\email{edgard@stanford.edu}
\affiliation{Department of Physics, Stanford University, Stanford, California 94305, USA}

\author{Frederi Viens}
\email{viens@msu.edu}
\affiliation{Department of Statistics and Probability, Michigan State University, East Lansing, Michigan 48824, USA}
\affiliation{Rice University, Department of Statistics, Houston, Texas 77005, USA}

\author{Jorge Piekarewicz}
\email{jpiekarewicz@fsu.edu}
\affiliation{Department of Physics, Florida State University, Tallahassee, FL 32306 USA}

\date{\today}

\begin{abstract}

A covariant energy density functional is calibrated using a principled Bayesian statistical framework 
informed by experimental binding energies and charge radii of several magic and semi-magic nuclei.
The Bayesian sampling required for the calibration is enabled by the emulation of the high-fidelity 
model through the implementation of a reduced basis method (RBM)---a set of dimensionality reduction 
techniques that can speed up demanding calculations involving partial differential equations by several 
orders of magnitude. The RBM emulator we build---using only 100 evaluations of the high-fidelity model---is 
able to accurately reproduce the model calculations in tens of milliseconds on a personal computer, an
increase in speed of nearly a  
factor of 3,300 when compared to the original solver. Besides the analysis of the posterior distribution 
of parameters, we present predictions with properly estimated uncertainties for 
observables not included in the fit, specifically the neutron skin thickness of $^{208}$Pb\,\cite{Abrahamyan:2012gp,
Horowitz:2012tj,Adhikari:2021phr} and $^{48}$Ca \cite{CREX:2022kgg}, as reported by PREX and
CREX collaborations.
The straightforward implementation and outstanding performance of the RBM makes it an ideal tool for  
assisting the nuclear theory community in providing reliable estimates with properly quantified uncertainties 
of physical observables. Such uncertainty quantification tools will become essential given the expected 
abundance of data from the recently inaugurated and future experimental and observational facilities.
\end{abstract}
\maketitle

\section{Introduction}\label{eeee}

Nuclear science is undergoing a transformational change enabled by the commissioning of new 
experimental and observational facilities as well as dramatic advances in high-performance 
computing\,\cite{LongRangePlan}. The newly operational Facility for Rare Isotope Beams (FRIB), 
together with other state-of-the-art facilities throughout the world, will produce short-lived isotopes 
that provide vital information on the creation of the heavy elements in the cosmos. In turn, earth and
space-based telescopes operating across the entire electromagnetic spectrum will constrain 
the nuclear dynamics in regimes inaccessible in terrestrial laboratories. Finally, improved and 
future gravitational-wave detectors will provide valuable insights into the production sites of the 
heavy elements as well as on the properties of ultra-dense matter at both low and finite 
temperatures\,\cite{Abbott:PRL2017,Drout:2017ijr,Cowperthwaite:2017dyu,Chornock:2017sdf,
Nicholl:2017ahq,Fattoyev:2017jql,Annala:2017llu,Abbott:2018exr}.

To fully capitalize on the upcoming discoveries, a strong synergy will need to be further developed between 
theory, experiment, and observation. First, theory is needed to decode the wealth of information 
contained in the new experimental and observational data. Second, new measurements drive 
new theoretical advances which, in turn, uncover new questions that motivate new experiments. 
From the theoretical perspective, there are a few frameworks that could be used to make sense of the experimental and observational data, each with their own respective advantages and drawbacks. One such model is found by adopting a nuclear interaction 
rooted in chiral effective field theory (EFT). Chiral  EFT --- a theoretical framework inspired by the 
underlying symmetries of QCD --- provides a systematic and improvable expansion in terms of a 
suitable small parameter, defined as the ratio of the length scale of interest to the length scale of 
the underlying  dynamics\,\cite{Weinberg:1990rz,vanKolck:1994yi,Ordonez:1995rz}. During the last 
decade, enormous progress has been made in our understanding of the equation of state (EOS) of 
pure neutron matter by systematically improving the chiral expansion\,\cite{Hebeler:2009iv,Tews:2012fj,
Kruger:2013kua,Lonardoni:2019ypg,Drischler:2021kxf,Sammarruca:2021mhv,Sammarruca:2022ser}. 
However, the chiral expansion breaks down once the relevant energy scale of the 
problem becomes comparable to the hard scale associated with the underlying dynamics.
This fact alone precludes the use of chiral perturbation theory in the study of high density 
matter. 

A more phenomenological approach that could be extended to higher densities is 
Density Functional Theory (DFT). Developed in quantum chemistry\,\cite{Kohn:1999}
but now widely used in nuclear physics, DFT is a powerful technique whose greatest virtue is 
shifting the focus away from the complicated many-body wave function that depends on the 
spatial coordinates of all particles, to an energy density functional (EDF) that depends only on 
the three spatial coordinates of the ground state density. Moreover, DFT guarantees that both 
the exact ground-state density and energy of the complicated many-body system may be 
obtained from minimizing a suitable functional \cite{Hohenberg:1964zz,Kohn:1965}. In an effort to simplify the solution of the problem, the 
Kohn-Sham formalism reformulates the DFT problem in favor of one-particle orbitals that may 
be obtained by solving self-consistently a set of equations that closely resemble the structure 
of the well-known Hartree equations\,\cite{Kohn:1965}. It is important 
to note that the theorems behind DFT offer no guidance on how to construct the correct EDF. 
This fact is mitigated in nuclear physics by incorporating as many physical insights as possible 
into the construction of the functional, and then calibrating the parameters of the model by using the available experimental and observational data.

The calibrated models are not static, however, and theory must be nimble in its response to the exciting new data that will emerge from 
future experiments and observations. In the particular case of DFT, new data must be
promptly incorporated into the refinement of the EDF to explore the full impact of the
new information. This is particularly relevant given that nuclear physics has the ability to 
predict the structure and properties of matter in regions inaccessible to either experiment 
or observation. For example, one may use Bayesian inference to identify strong correlations
between a desired property, which cannot be measured, and a surrogate observable 
that may be determined experimentally. However, Bayesian methods often require multiple 
evaluations of the same set of observables for many different realizations of the model parameters.
If the nuclear observables informing the EDF are computationally expensive, then direct Bayesian
inference is highly impractical. This computational challenge has motivated many of the recent efforts by the nuclear theory community in the development and adoption of emulators to accelerate computation speed with a minimal precision loss \cite{phillips2021get, frame2018eigenvector,konig2020eigenvector,furnstahl2020efficient,melendez2021fast,drischler2021toward, Bonilla:2022rph,sarkar2022self,boehnlein2022colloquium,tews2022nuclear,higdon2015bayesian,mcdonnell2015uncertainty,anderson2022applications}.  In this work we explore the application of one such class of emulators, the Reduced Basis Method (RBM)~\cite{quarteroni2015reduced,hesthaven2016certified}, which falls under the umbrella of the general Reduced Order Models (ROM) techniques
~\cite{quarteroni2014reduced,brunton2019data}.

The Reduced Basis Method encapsulates a set of dimensionality reduction approaches that generally aim at speeding up computations by approximating the solution to differential equations with just a handful of active components (the reduced basis). These methods have been shown to exhibit speed increases of several orders of magnitude in various areas of science and engineering  \cite{benner2015survey,sartori2016reduced,quarteroni2011certified,field2011reduced}, including specific applications for uncertainty quantification~\cite{nguyen2010reduced,jiang2016goal}, and have been recently demonstrated to be viable for applications in nuclear physics DFT\,\cite{Bonilla:2022rph}. Solving the full system
of differential equations self-consistently in the framework of covariant DFT is not a particularly demanding computational task for modern computers, usually taking
around a minute
for a heavy nucleus such as ${}^{208}$Pb.
The computational bottleneck appears when millions of such evaluations must be carried out sequentially to perform Bayesian inference, and the problem multiplies when several nuclei are involved or if one wants to consider and compare different EDFs. A speed-up factor of three orders of magnitude or more provided by techniques such as the RBM could bridge the computational gap and enable new scientific advancements that would otherwise be impossibly or significantly more expensive. The adoption of these methods, paired together with leadership-class computing infrastructure, will enable the quick response that is needed to take full advantage of the vast wealth of experimental and observational data that will be coming in the next years.

The intent of this manuscript is to develop and showcase a pipeline for the calibration and uncertainty quantification of a nuclear model --- a covariant energy density functional --- enabled by the RBM emulation. To that goal, in Sec.~\ref{Sec:RMF} we provide a brief introduction to the relativistic mean field model we employ, culminating with the set of differential equations that need to be solved in order to
calculate nuclear observables. In Sec.~\ref{Sec: RBM} we present the reduced basis methodology, alongside an explanation on how it is used to construct an emulator that simplifies the computations of the DFT model. In Sec.~\ref{Sec: Bayesian} we explain the theory and implementation of the Bayesian statistical analysis used to calibrate the model parameters, with full uncertainty quantification. In Sec.~\ref{Sec: Results} we present and discuss the results of the calibration, displaying
the Bayesian posterior distribution of the model parameters, together with the model predictions with quantified uncertainties for the neutron skin thickness of both  $^{208}$Pb and $^{48}$Ca. These observables have been the focus of recent experimental campaigns~\cite{Abrahamyan:2012gp,Adhikari:2021phr,CREX:2022kgg}, and its widespread implications are of great interest to the nuclear physics and astrophysics communities~\cite{Reed:2021nqk,Reinhard:2022inh}  
Finally, in Sec.~\ref{Sec: Conclusions and outlook} we present our conclusions and outlooks, with a perspective on the role that this class of emulators could play, in the near future, on the nuclear theory-experiment cycle enhanced by statistics and machine learning \cite{ireland2015enhancing,boehnlein2022colloquium,bedaque2021ai}.

\section{Relativistic Mean Field Calculations} \label{Sec:RMF}

The cornerstone of covariant density functional theory is a Lagrangian density that includes 
nucleons and mesons as the effective degrees of freedom. Besides the photon that mediates 
the long-range Coulomb interaction, the model includes the isoscalar-scalar $\sigma$ meson, 
the isoscalar-vector $\omega$ meson, and the isovector-vector $\rho$ 
meson\,\cite{Walecka:1974qa,Serot:1984ey}. The interacting Lagrangian density consists of a
nucleon-nucleon interaction mediated by the various mesons alongside non-linear meson 
interactions\,\cite{Boguta:1977xi,Serot:1984ey,Mueller:1996pm,Horowitz:2000xj,Todd:2003xs,
Chen:2014sca}. That is,
\begin{widetext}
\begin{eqnarray}
{\mathscr L}_{\rm int} &=&
\bar\psi \left[g_{\rm s}\phi   \!-\! 
         \left(g_{\rm v}V_\mu  \!+\!
    \frac{g_{\rho}}{2}{\mbox{\boldmath$\tau$}}\cdot{\bf b}_{\mu} 
                               \!+\!    
    \frac{e}{2}(1\!+\!\tau_{3})A_{\mu}\right)\gamma^{\mu}
         \right]\psi \nonumber \\
                   &-& 
    \frac{\kappa}{3!} (g_{\rm s}\phi)^3 \!-\!
    \frac{\lambda}{4!}(g_{\rm s}\phi)^4 \!+\!
    \frac{\zeta}{4!}   g_{\rm v}^4(V_{\mu}V^\mu)^2 +
   \Lambda_{\rm v}\Big(g_{\rho}^{2}\,{\bf b}_{\mu}\cdot{\bf b}^{\mu}\Big)
                           \Big(g_{\rm v}^{2}V_{\nu}V^{\nu}\Big).                                           
 \label{LDensity}
\end{eqnarray}
\end{widetext}
The first line in the above expression includes Yukawa couplings of the meson fields 
to the appropriate bilinear combination of nucleon fields. In turn, the second line 
includes non-linear meson interactions that serve to simulate the complicated many-body 
dynamics and  that are required to improve the predictive power of the 
model\,\cite{Boguta:1977xi,Mueller:1996pm,Horowitz:2000xj}. For a detailed account 
on the physics underlying each terms in the Lagrangian see Refs.\,\cite{Chen:2014sca,
Yang:2019fvs}.

\subsection{Meson Field Equations}

In the mean-field limit, both the meson-field operators and their corresponding sources
are replaced by their ground state expectation values. For spherically symmetric systems, 
all meson fields and the photon satisfy Klein-Gordon equations of the following 
form\,\cite{Todd:2003xs}:
\begin{widetext}
\begin{subequations}
\begin{eqnarray}
 &&
 \left(
  \frac{d^{2}}{dr^{2}}+\frac{2}{r}\frac{d}{dr}-m_{\rm s}^{2}
 \right)\Phi_{0}(r) - g_{\rm s}^{2}
 \left(
  \frac{\kappa}{2}\Phi_{0}^{2}(r)  +
  \frac{\lambda}{6}\Phi_{0}^{3}(r) 
 \right)=-g_{\rm s}^{2}
 \Big(\rho_{\rm s,p}(r)+\rho_{\rm s,n}(r)\Big), \label{KGEqns: Phi} \\
 &&
 \left(
  \frac{d^{2}}{dr^{2}}+\frac{2}{r}\frac{d}{dr}-m_{\rm v}^{2}
 \right)W_{0}(r) -  g_{\rm v}^{2}
 \left(
  \frac{\zeta}{6}W_{0}^{3}(r) + 
  2\Lambda_{\rm v}B_{0}^{2}(r)W_{0}(r)
 \right)=-g_{\rm v}^{2} \Big(\rho_{\rm v,p}(r)+\rho_{\rm v,n}(r)\Big), \label{KGEqns: W} \\
 &&
 \left(
  \frac{d^{2}}{dr^{2}}+\frac{2}{r}\frac{d}{dr}-m_{\rho}^{2}
 \right)B_{0}(r) - 2\Lambda_{\rm v}g_{\rho}^{2}W_{0}^{2}(r)B_{0}(r)
 =-\frac{g_{\rho}^{2}}{2}\Big(\rho_{\rm v,p}(r)-\rho_{\rm v,n}(r)\Big), \label{KGEqns: B} \\
 &&
 \left(
  \frac{d^{2}}{dr^{2}}+\frac{2}{r}\frac{d}{dr}\right)A_{0}(r)  
  =-e\rho_{\rm v,p}(r), \label{KGEqns: A}
\end{eqnarray}
\label{KGEqns}
\end{subequations}
\end{widetext}
where we have defined $\Phi\!=\!g_{\rm s}\phi$, $W_{\mu}\!=\!g_{\rm v}V_{\mu}$, 
and ${\bf B}_{\mu}\!=\!g_{\rho}{\bf b}_{\mu}$. The various meson masses, which are 
inversely proportional to the effective range of the corresponding meson-mediated 
interaction, are given by $m_{\rm s}$, $m_{\rm v}$, and $m_{\rho}$. The 
source terms for the Klein-Gordon equations are ground-state densities with 
the correct Lorentz and isospin structure. Finally, the above scalar (${\rm s}$) and 
vector (${\rm v}$) densities are written in terms of the occupied proton and neutron
Dirac orbitals: 
\begin{equation}
  \left(
    \begin{array}{c}
      \rho_{\rm s,t}(r) \\
      \rho_{\rm v,t}(r) 
    \end{array}	
   \right) = \sum_{n\kappa}^{\rm occ}
   \left(\frac{2j_{\kappa}+1}{4\pi r^{2}}\right)
   \Big(g_{n\kappa t}^{2}(r)\mp f_{n\kappa t}^{2}(r)\Big). 
  \label{RhoSV} 
\end{equation}
Here $t$ identifies the nucleon species (or isospin) and $n$ denotes the principal 
quantum number. We note that some of the semi-magic nuclei that will be used to 
calibrate the energy density functional may have open protons or neutron shells. In such 
case, we continue to assume spherical symmetry, but introduce a fractional occupancy 
for the valence shell. For example, in the particular case of ${}^{116}$Sn, only two 
neutrons occupy the valence $d_{3/2}$ orbital, so the filling fraction is set to $1/2$.

\subsection{Dirac Equations}

In turn, the nucleons satisfy a Dirac equation with scalar and time-like vector 
potentials generated by the meson fields. The eigenstates of the Dirac equation 
for the spherically symmetric ground state assumed here may be classified 
according to a generalized angular momentum $\kappa$. The orbital angular 
momentum $l$ and total angular momentum $j$ are obtained from $\kappa$
as follows:
\begin{equation}
 j = |\kappa| - \frac {1}{2} \;; \quad
 l = \begin{cases}
               \kappa\;,  & {\rm if} \; \kappa>0; \\
        -(1+\kappa)\;,  & {\rm if} \; \kappa<0, 
     \end{cases}
\end{equation}

where $\kappa$ takes all integer values different from zero. For example, 
$\kappa\!=\!-\!1$ corresponds to the $s_{1\!/2}$ orbital. The 
single-particle solutions of the Dirac equation may then be written as 
\begin{equation}
 {\cal U}_{n \kappa m t}({\bf r}) = \frac {1}{r}
 \left( \begin{array}{c}
   \phantom{i}
   g_{n \kappa t}(r) {\cal Y}_{+\kappa \mbox{} m}(\hat{\bf r})  \\
  if_{n \kappa t}(r) {\cal Y}_{-\kappa \mbox{} m}(\hat{\bf r})
 \end{array} \right),
\label{Uspinor}
\end{equation}
where $m$ is the magnetic quantum number and the spin-spherical harmonics 
${\cal Y}_{\kappa \mbox{} m}$ are obtained by coupling the orbital angular $l$
momentum and the intrinsic nucleon spin to a total angular momentum
$j$. However, note that the orbital angular momentum of the upper and lower 
components differ by one unit, indicating that the orbital angular momentum is 
not a good quantum number. The functions $g_{n \kappa t}$ and $f_{n \kappa t}$ 
satisfy a set of first order, coupled differential 
equations that must be solved to obtain the single particle spectrum:
\begin{widetext}
\begin{subequations}
\begin{eqnarray}
  &&
  \left(\frac{d}{dr}+\frac{\kappa}{r}\right)g_{a}(r)
 -\left[E_{a}+M-\Phi_{0}(r)-W_{0}(r)\mp\frac{1}{2}B_{0}(r)
 -e\left\{
    \begin{array}{c}  1 \\  0 \end{array}
  \right\}A_{0}(r)
  \right]f_{a}(r)=0\;, \phantom{xxxx} \label{Eq: Dirac g} \\
  &&
  \left(\frac{d}{dr}-\frac{\kappa}{r}\right)f_{a}(r)
 +\left[E_{a}-M+\Phi_{0}(r)-W_{0}(r)\mp\frac{1}{2}B_{0}(r)
 -e\left\{
    \begin{array}{c}  1 \\  0 \end{array}
  \right\}A_{0}(r)
  \right]g_{a}(r)=0\;,  \label{Eq: Dirac f}
\end{eqnarray}
\label{DiracEqn}
\end{subequations}
\end{widetext}
where the upper numbers correspond to protons and the lower ones to neutrons, and we have
used the shorthand notation $a\!=\!\{n\kappa t\}$ to denote the relevant quantum numbers.
Finally, $g_{a}(r)$ and $f_{a}(r)$ satisfy the following normalization condition:
\begin{equation}
    \int_{0}^{\infty} \Big(g_a^2(r) +f_a^2(r) \Big) dr=1.
 \label{Eq: norm}
\end{equation}
Looking back at Eq.(\ref{RhoSV}), we observe that the proton and neutron vector densities are 
conserved, namely, their corresponding integrals yield the number of protons $Z$ and the 
number of neutrons $N$, respectively. In contrast, the scalar density is not conserved.

\subsection{Ground state properties}

From the solution of both the Klein-Gordon equations for the mesons~\eqref{KGEqns} and the Dirac equation for the nucleons~\eqref{DiracEqn}, we can calculate all ground-state properties of a nucleus composed of $Z$ protons and $N$ neutrons. The proton and neutron mean square radii are determined directly in terms of their respective vector densities:
\begin{subequations}
\begin{eqnarray}
    R_p^2 \equiv \frac{4\pi}{Z} \int_{0}^{\infty} r^4 \rho_\text{v,p}(r) dr, \\
    R_n^2 \equiv \frac{4\pi}{N} \int_{0}^{\infty} r^4 \rho_\text{v,n}(r) dr.
\end{eqnarray}
\label{Eq: Radius}
\end{subequations}

Following \cite{Chen:2014sca} we approximate the charge radius of the nucleus by folding the finite size of the proton $r_p$ as:

\begin{equation}
    R_\text{ch}^2 = R_p^2 + r_p^2,
    \label{Eq: Charge Rad}
\end{equation}

where we have used for the radius of a single proton $r_p=0.84$ fm \cite{tiesinga2021codata}. 

In turn, the total binding energy per nucleon $E/A-M$, includes contributions from both the nucleon and meson fields: $E\!=\!E_\text{nuc}\!+\!E_\text{mesons}$. The nucleon contribution 
is calculated directly in terms of the single particle energies obtained from the solution of the Dirac equation. That is,
\begin{equation}
    E_\text{nuc} = \sum_a^\text{occ} (2j_{a}+1)\,E_a,
    \label{Eq: Nuc En}
\end{equation}
where the sum is over all occupied single particle orbitals, $E_{a}$ is the energy of the $a_{\rm th}$ orbital, and $(2j_{a}+1)$ is the maximum occupancy of such orbital. For partially filled shells, one must multiply the above expression by the corresponding filling fraction, which in the case of $^{116}$Sn, is equal to $1$ for all orbitals except for the valence $d_{3/2}$ neutron orbital where the filling fraction is $1/2$.

The contribution to the energy from the meson fields and the photon may be written as
\begin{equation}
    E_\text{mesons}=4\pi\int_{0}^{\infty}\!\!\Big(E_{\sigma} + E_{\omega} +E_{\rho} + E_{\gamma}  + E_{\omega\rho}\Big) r^2\,dr,
    \label{Eq: En Fields}
\end{equation}
where the above expression includes individual contributions from the $\sigma$, $\omega$, and $\rho$ mesons, the photon, and the mixed $\omega\rho$ term. In terms of the various meson fields and the ground-state nucleon densities, the above contributions are given by

\begin{widetext}
\begin{subequations}
\begin{eqnarray}
 && E_{\sigma}  = \frac{1}{2}\Phi_0(r)\Big(\rho_{\rm s,p}(r)+\rho_{\rm s,n}(r)\Big) - \frac{\kappa}{12}\Phi^3_0(r) - \frac{\lambda}{24}\Phi^4_0(r) , \\
&& E_{\omega} =-\frac{1}{2}W_0(r)\Big(\rho_{\rm v,p}(r)+\rho_{\rm v,n}(r)\Big) + \frac{\zeta}{24} W^4_0(r) , \\
&& E_{\rho} = -\frac{1}{4}B_0(r)\Big(\rho_{\rm v,p}(r)-\rho_{\rm v,n}(r)\Big) , \\
&& E_{\gamma} =-\frac{1}{2}eA_0(r)\rho_{\rm v,p}(r) , \\
&& E_{\omega\rho} =\Lambda_\text{v} W_0^2(r)B_0^2(r) .
\end{eqnarray}
\label{Eq: Field Energy}
\end{subequations}
\end{widetext}

Following \cite{Chen:2014sca}, in this work we calibrate the relativistic mean field model by comparing the calculations of charge radii and binding energies with the experimentally measured values for the doubly magic and semi-magic nuclei: $^{16}$O, $^{40}$Ca, $^{48}$Ca, $^{68}$Ni\footnote{Both the charge radius of $^{68}$Ni and $^{100}$Sn have not been measured yet. Therefore, our dataset consists of 18 points, 10 binding energies and 8 charge radii.}, $^{90}$Zr, $^{100}$Sn, $^{116}$Sn, $^{132}$Sn, $^{144}$Sm, $^{208}$Pb.

\subsection{Bulk properties parametrization}

The Lagrangian density of Eq.~\eqref{LDensity} is defined in terms of seven coupling constants. These 
seven parameters plus the mass of the $\sigma$ meson define the entire 8-dimensional parameter 
space (the masses of the two vector mesons are fixed at their respective experimental values of
$m_{\rm v}\!=\!782.5\,{\rm MeV}$ and $m_{\rho}=\!763\,{\rm MeV}$). Once the theoretical model 
and the set of physical observables informing the calibration have been specified, one proceeds to
sample the space of model parameters: $\alpha\!\equiv\!\{m_{\rm s}, g_{\rm s}, g_{\rm v},g_{\rho},
\kappa,\lambda,\Lambda_{\rm v},\zeta\}$. However, given that the connection between the model 
parameters and our physical intuition is tenuous at best, the sampling algorithm can end up wandering 
aimlessly through the parameter space. The problem is further exacerbated in covariant DFT by the 
fact that the coupling constants are particularly large. Indeed, one of the hallmarks of the covariant
framework is the presence of strong---and cancelling---scalar and vector potentials. So, if the scalar 
coupling $g_{\rm s}$ is modified without a compensating modification to the vector coupling 
$g_{\rm v}$, it is likely that no bound states will be found. To overcome this situation 
one should make correlated changes in the model parameters. Such correlated changes can be
implemented by taking advantage of the fact that some of the model parameters can 
be expressed in terms of a few bulk properties of infinite nuclear 
matter\,\cite{Glendenning:2000,Chen:2014sca}. Thus, rather than sampling the model parameters
$\alpha$, we sample the equivalent bulk parameters 
$\theta\!=\!\{m_{\rm s},\rho_{0},\epsilon_{0},M^{\ast},K,J,L,\zeta\}$. In this expression, $\rho_{0}$,
$\epsilon_{0}$, $M^{\ast}$, and $K$ are the saturation density, the binding energy, effective 
nucleon mass, and incompressibility coefficient of symmetric nuclear matter evaluated at saturation 
density. In turn, $J$ and $L$ are the value and slope of the symmetry energy also at saturation 
density. The quartic vector coupling $\zeta$ is left as a ``bulk" parameter as the properties of infinite nuclear 
matter at saturation density are largely insensitive to the value of $\zeta$\,\cite{Mueller:1996pm}. 
The virtue of such a transformation is twofold: first, most of the bulk parameters given in $\theta$ are
already known within a fairly narrow range, making the incorporation of Bayesian priors easier and natural, and second, a modification to the bulk parameters involves 
a correlated change in several of the model parameters, thereby facilitating the success of the calibration
procedure.

In the context of density functional theory, Eqs.(\ref{KGEqns}-\ref{DiracEqn}) represent the effective 
Kohn-Sham equations for the nuclear many-body problem. Once the Lagrangian parameters $\alpha$ have been calculated from the chosen bulk parameters $\theta$, these set of non-linear coupled equations  must be solved self-consistently. That is, the single-particle orbitals satisfying the Dirac equation are 
generated from the various meson fields which, in turn, satisfy Klein-Gordon equations with the appropriate 
ground-state densities as the source terms. This demands an iterative procedure in which mean-field 
potentials of the Wood-Saxon form are initially provided to solve the Dirac equation for the occupied 
nucleon orbitals which are then combined to generate the appropriate densities for the meson field. 
The Klein-Gordon equations are then solved with the resulting meson fields providing a refinement 
to the initial mean-field potentials. This procedure continues until self-consistency is achieved, namely,
the iterative process has converged.

In the next section we show how the reduced basis method bypasses such a complex and time-consuming 
procedure by constructing suitable reduced bases for the solution of both the Klein-Gordon and Dirac 
equations.

\section{The Reduced Basis Method}\label{Sec: RBM}

A system of coupled differential equations in the one-dimensional variable $r$, such as Eqs.~\eqref{KGEqns} and \eqref{DiracEqn}, can be computationally solved by numerical methods such as finite element or Runge-Kutta. We shall refer to the numerical solutions obtained from those computations as high fidelity solutions for the rest of the discussion. Those approaches possess an intrinsic resolution $\Delta r$---such as the grid spacing in the case of finite element or the step size in the case of Runge-Kutta\footnote{Both the grid size and the step size could be adaptive instead of constant across the spatial domain. We shall assume a constant $\Delta r$ for the rest of the discussion for the sake of simplicity.}. For a given interval $L$ in which we are interested in solving the equations, each of the functions involved will have roughly $\mathcal{N}\sim \frac{L}{\Delta r}$ elements. In the case of the finite element method for example, for fixed particle densities and a given grid, the four fields involved in Eqs.~\eqref{KGEqns} become arrays of unknown values:
\begin{equation}
\begin{split}
    & r\rightarrow [r_1,\ r_2\ , ... \ , \ r_\mathcal{N}], \\
    & \Phi_0(r)\rightarrow [\Phi_0(r_1),\ \Phi_0(r_2)\ , ... \ , \ \Phi_0(r_\mathcal{N})], \\
    & W_0(r)\rightarrow [W_0(r_1),\ W_0(r_2)\ , ... \ , \ W_0(r_\mathcal{N})], \\
    & B_0(r)\rightarrow [B_0(r_1),\ B_0(r_2)\ , ... \ , \ B_0(r_\mathcal{N})],\\
    & A_0(r)\rightarrow [A_0(r_1),\ A_0(r_2)\ , ... \ , \ A_0(r_\mathcal{N})].
\end{split}
\label{Eq: VectorEqs}
\end{equation}

In turn, once the differential operators such as $\frac{d^2}{dr^2}$ are transformed into matrices of finite differences, the differential equations themselves will become matrix equations for the unknown arrays \eqref{Eq: VectorEqs}. The same procedure follows for the Dirac equations~\eqref{DiracEqn} for fixed fields, with each upper and lower components $g_{n\kappa}(r)$ and $f_{n\kappa}(r)$ for protons and neutrons becoming arrays of unknown values that must be solved for.

Both the traditional Runge-Kutta solver 
and the finite element solver we developed to iteratively tackle Eqs~\eqref{KGEqns} and \eqref{DiracEqn} have $L=20$ fm, $\Delta r= 0.05$ fm, and therefore $\mathcal{N}=400$.
The goal of the Reduced Basis (RB) approach is to build a framework that, after a preparation period called the offline stage, can obtain approximate solutions to the differential equations with as few---or even better, none---calculations of size $\mathcal{N}$ during the evaluation period called the online stage \cite{hesthaven2016certified}. Any observable computed from such solutions, such as binding energies and radii, should also involve as few $\mathcal{N}$ calculations as possible to streamline the uncertainty quantification procedure.

The RBM implementation we construct in this work consists of two principal steps: ``training and projecting" \cite{Bonilla:2022rph}. In the  first step we build the RB using information from high fidelity solutions, while in the second step we create the necessary equations for finding the approximate solution by projecting over a well-chosen low-dimensional subspace. The following subsections explain both steps in detail.

\subsection{Training}

\begin{figure}[h]
    \centering
    \includegraphics[width=0.48\textwidth]{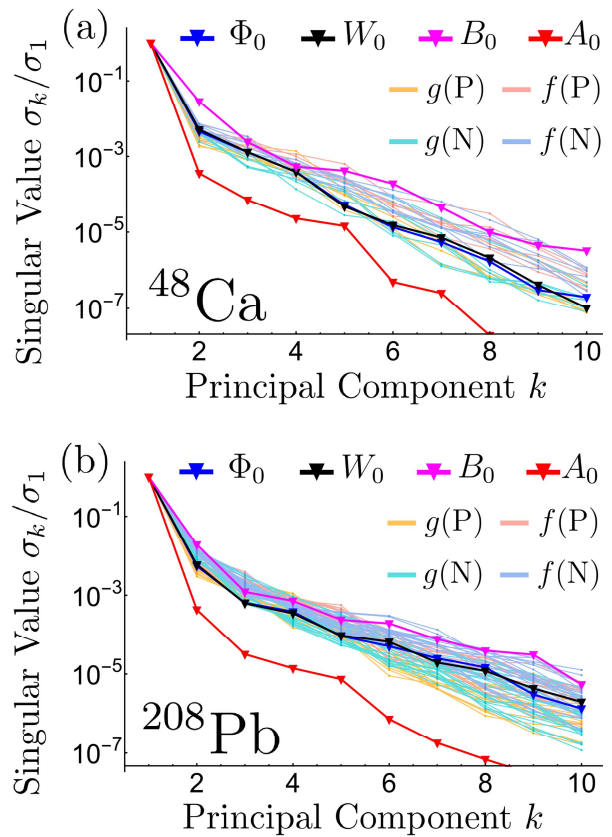}
    \caption{Normalized singular values $\sigma_k/\sigma_1$ for the fields $\Phi_0(r)$, $W_0(r)$, $B_0(r)$, and $A_0(r)$, and the single particle wave functions of the upper $g_{n\kappa}(r)$ and lower $f_{n\kappa}(r)$ components for $^{48}$Ca (a) and $^{208}$Pb (b). The single particle proton levels are denoted as $g$(P) and $f$(P) for the upper and lower components, respectively, while the single particle neutron levels are denoted as $g$(N) and $f$(N) for the upper and lower components, respectively. There are six proton levels and seven neutron levels for $^{48}$Ca (a), while for $^{208}$Pb there are sixteen proton levels and twenty-two neutron levels.}
        \label{fig: PrincipalComponents}
\end{figure}

We begin by proposing the corresponding RB expansion for each function involved in Eqs.~\eqref{KGEqns} and~\eqref{DiracEqn}:
\begin{subequations}
\begin{eqnarray}
 &&
 \Phi_0(r) \approx \hat \Phi_0(r) = \sum_{k=1}^{n_{\Phi}} a_k^{\Phi}\  \Phi_k(r), \\
 &&
 W_0(r) \approx \hat W_0(r) = \sum_{k=1}^{n_{W}} a_k^{W}\  W_k(r), \\
 &&
 B_0(r) \approx \hat B_0(r) = \sum_{k=1}^{n_{B}} a_k^{B}\  B_k(r), \\
 &&
 A_0(r) \approx \hat A_0(r) = \sum_{k=1}^{n_{A}} a_k^{A}\  A_k(r), \\
 &&
g(r) \approx \hat g(r) = \sum_{k=1}^{n_{g}} a_k^{g}\  g_k (r), \\
 &&
f(r) \approx \hat f(r) = \sum_{k=1}^{n_{f}} a_k^{f}\  f_k(r),
\end{eqnarray}
\label{Eq: RB}
\end{subequations}

The subscripts $n$ and $\kappa$ have been omitted from the $g_{n\kappa}$ and $f_{n\kappa}$ components for the sake of clarity, but it is important to note that the expansion will have unique coefficients $a_k$, and possible different number of basis $n_g$ and $n_f$  for each level. The functions with sub-index $k$, $A_k(r)$ for example, form the RB used to build their respective approximations, $\hat A_0(r)$ in this case. It is interesting to note that Eq.~\eqref{KGEqns: A} can be solved to explicitly obtain $A_0(r)$ as integrals of the proton density (see Eq.(7) in \cite{Todd:2003xs}). Nevertheless, we found that expanding $A_0(r)$ in its own RB resulted in appreciably bigger speed up gains by the RBM emulator with negligible loss in accuracy.

\begin{figure*}[t]
    \centering
    \includegraphics[width=0.95\textwidth]{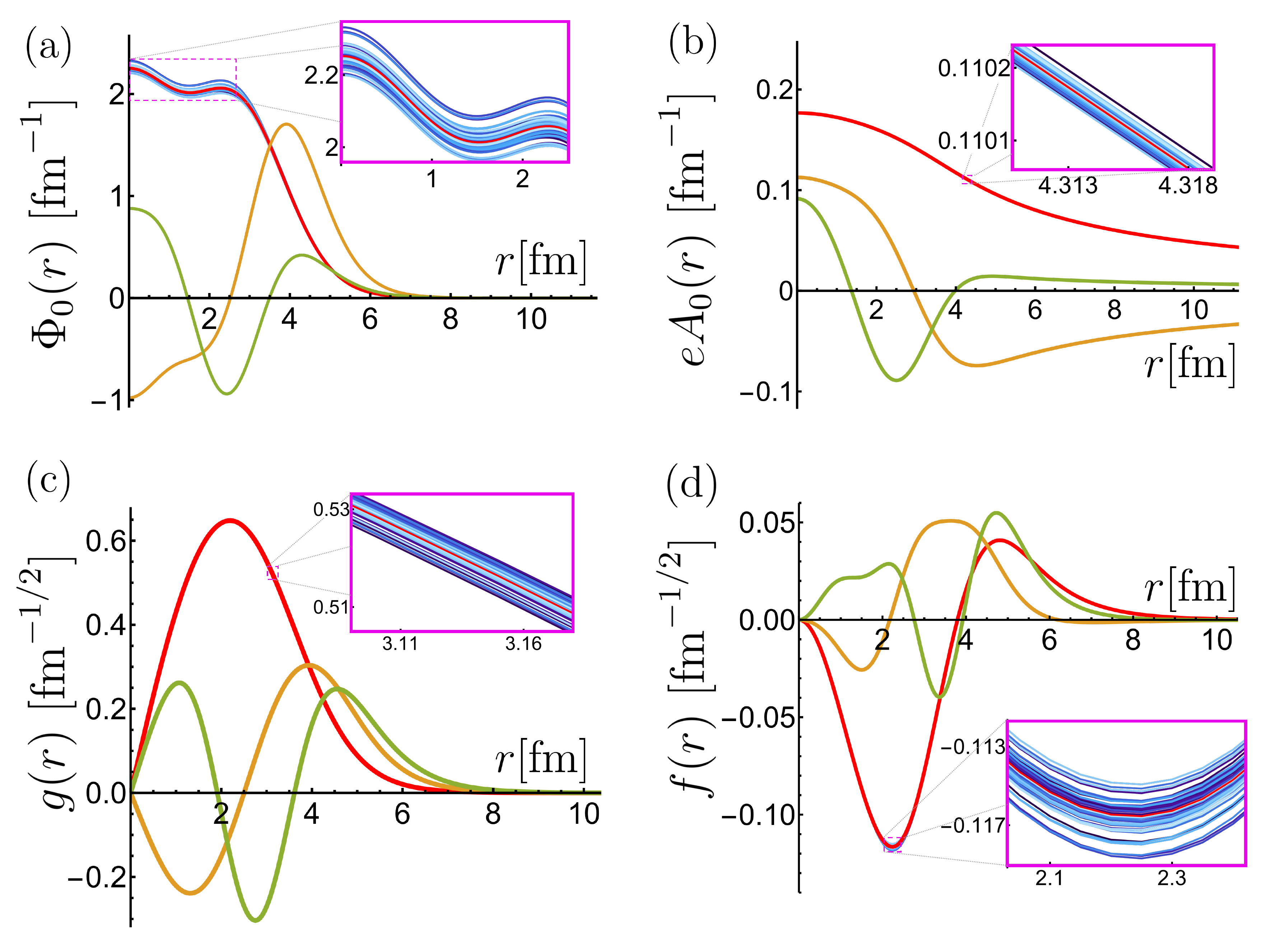}
    \caption{
    First three principal components in red, orange, and green respectively, for $\Phi_0(r)$ (a), $eA_0(r)$ (b), the first $g(r)$ for neutrons (c), and the sixth $f(r)$ for protons in $^{48}$Ca. The second and third components (in orange and green) have been re-scaled arbitrarily for plotting convenience. The 50 solutions used in the training set are shown in different shades of blue in each figure. The spread of such solutions is barely visible for $\Phi_0(r)$ and $f(r)$, and almost undetectable for the other two cases. The spread is further enhanced in the inset plots within the magenta squares in each sub figure.}
    \label{fig:CaPCA}
\end{figure*}

Once chosen, each RB is fixed and will not change when finding approximated solutions to Eqs.~\eqref{KGEqns} and~\eqref{DiracEqn} for different parameters $\alpha$. The coefficients $a_k^{(\cdot)}$ do depend on the parameters $\alpha$ and are the ones responsible for adjusting the approximate solution as the parameter space is explored. It is important to note that, if there is a level crossing, the occupancy configuration of the nucleus will change. The RBM implementation we describe here---relying on smooth variations of the functions involved as $\alpha$ changes---is unable to correctly emulate the solution if suddenly an orbital looses or gains nucleons. For the parameter ranges we studied we don't expect that to happen for the closed shell magic nuclei we employed given the gap in the single particle spectrum\footnote{For future applications involving, for example, nuclear deformations it will be important to modify the RB accordingly.}. We did not observe level crossing either on the partially filled neutrons and protons shells of $^{116}$Sn and $^{144}$Sm.

There are several approaches for the construction of the reduced basis \cite{hesthaven2016certified,quarteroni2015reduced}, most of which involve information obtained from high fidelity solutions for a sample of the parameters $\alpha$. For this work, we choose the Proper Orthogonal Decomposition (POD) approach, which consists of building the RB as the first $n$ components (singular vectors) of a Singular Value Decomposition (SVD) \cite{blum_hopcroft_kannan_2020}---see also Principal Component Analysis (PCA) \cite{jolliffe2002principal}---performed on a set of high fidelity solutions.

For each nucleus involved we compute high fidelity evaluations for 50 parameter samples obtained from the calibration performed in \cite{Chen:2014sca}. We perform the SVD on each of the four fields and wave functions for the respective protons and neutron levels for all ten nuclei considered in this study. Figure~\ref{fig: PrincipalComponents} shows the normalized singular values $\sigma_k/\sigma_1$ for the field and nucleon wave functions for $^{48}$Ca and $^{208}$Pb. Each singular value represents how much of the total variance in the entire sample that particular component is capable of explain~\cite{brunton2019data}. A fast exponential decay of the singular values can indicate that a RB with few components should be able to approximate the full solution with good accuracy (see also the discussion on the Kolmogorov $N-$width in Chapter V of \cite{quarteroni2015reduced}).

Figure~\ref{fig:CaPCA} shows the first three principal components obtained from the SVD of the 50 high fidelity evaluations for the $\Phi_0(r)$ and $A_0(r)$ fields, the upper component $g(r)$ of the first neutron level, and the lower component $f(r)$ for the last proton level for $^{48}$Ca. The figure also shows the corresponding 50 high fidelity solutions, although the spread is barely noticeable for the two wave function components, and is imperceptible outside of the inset plot for the photon field $A_0(r)$. We observed a similar small spread of the fields and wave functions for all the nuclei considered for the 50 high fidelity evaluations. This is consistent with the fact that the relativistic mean field model has been calibrated to reproduce ground state experimental observables such as masses and radii within a $0.5\%$ error\footnote{With an error of around $1.4\%$, the charge radius of $^{16}$O can be treated as an outlier in which the mean field approximation might break down.} \cite{Chen:2014sca}. Appreciable variations of the solutions would deteriorate such values. 

Choosing how many reduced bases to include for each field or wave function---the upper limits on the sums in Eqs.~\eqref{Eq: RB} $[n_\Phi,n_W,n_B,n_A,n_g,n_f]$---is a non-trivial process. In general, the more basis used the more precise the approximation will be, but that comes at the trade-off of an increased calculation time. This choice will not depend only on the relative importance of the singular values shown in Figure~\ref{fig: PrincipalComponents}, but rather on the quantities we are interested in calculating after solving the coupled differential equations. For example, for $^{48}$Ca, the photon field $A_0(r)$ has the fastest decaying singular values shown in Figure~\ref{fig: PrincipalComponents}, which could indicate that we need a smaller basis to reproduce it to the same level of accuracy than any of the other fields, such as $B_0(r)$ . Nevertheless, if our primary objective is to obtain accurate calculations for binding energies and charge radii, for example, it might be the case that we need to reproduce $A_0(r)$ to much better precision than $B_0(r)$, requiring $n_A>n_B$. We elaborate this discussion later when we describe our method for selecting the number of basis for each function. 

\subsection{Projecting}

For a fixed nucleus and a chosen RB configuration we have $n_\Phi+n_W+n_B+n_A$ free coefficients for the fields, $\sum_{i=1}^{l_P} \big(n_g^{(i,P)}+n_f^{(i,P)}\big)$ coefficients for the single particle wave functions for protons, and $\sum_{i=1}^{l_N} \big(n_g^{(i,N)}+n_f^{(i,N)}\big)$ for the single particle wave functions for neutrons. In these expressions $l_P$ and $l_N$ denote the total levels of protons and neutrons for the given nucleus, respectively. Additionally, since Eqs.~\eqref{DiracEqn} are eigenvalue equations, the respective energies $E_{i, P}$ and $E_{i, N}$ for each of the protons and neutrons levels also count as unknown quantities that need to be determined. Let us denote a list of such coefficients and energies as $\boldsymbol{a} \equiv \{a_1^\Phi,\ a_2^\Phi,\ . . . ,\ a_1^W, ... \ E_{1_P},\ E_{2_P}, ...  \}$.

For example, consider we are working with $^{48}$Ca which has six proton levels and seven neutron levels. If we set 3 RB for every field and wave function expansion in Eqs.~\eqref{Eq: RB}, we will have 12 coefficients associated with the fields, 36 coefficients and 6 energies associated with the protons, and 42 coefficients and 7 energies associated with the neutrons. This amounts for a total of 103 unknown quantities that must be determined from 103 equations. Each single particle level for protons and neutrons has an associated normalization condition shown in Eq.~\eqref{Eq: norm}. These normalization equations go in par with the unknown energies. The rest of the unknown coefficients---90 in this example---are determined from the Galerkin projection equations that we now describe. The Galerkin method \cite{rawitscher2018galerkin} is the traditional approach for obtaining such coefficients in the RBM~\cite{hesthaven2016certified, quarteroni2014reduced}.

Let us denote the set of field functions and wave functions in the compact notation $\Xi\equiv \{\Phi_0, W_0, B_0, A_0, g, f \}$ and their respective RB approximation $\hat \Xi\equiv \{\hat \Phi_0, \hat W_0, \hat B_0, \hat A_0, \hat g, \hat f \}$. Let us denote the Klein-Gordon and Dirac equations as operators acting on the set $\Xi$, re-arrange them such that they are all equal to 0, and label them as: 
\begin{subequations}
\begin{eqnarray*}
 &&
  \text{Eq}.\eqref{KGEqns: Phi} \rightarrow F_\alpha^\Phi[\Xi]=0, \\
 &&
\text{Eq}.\eqref{KGEqns: W} \rightarrow F_\alpha^W[\Xi]=0, \\
 &&
  \text{Eq}.\eqref{KGEqns: B} \rightarrow F_\alpha^B[\Xi]=0, \\
 &&
  \text{Eq}.\eqref{KGEqns: A} \rightarrow F_\alpha^A[\Xi]=0, \\
 &&
  \text{Eq}.\eqref{Eq: Dirac g} \rightarrow F_\alpha^g[\Xi]=0, \\
 &&
  \text{Eq}.\eqref{Eq: Dirac f} \rightarrow F_\alpha^f[\Xi]=0.
\end{eqnarray*}
\label{Eq: relabel}
\end{subequations}

For example, $F_\alpha^A[\Xi]=0$ reads $\left(\frac{d^{2}}{dr^{2}}+\frac{2}{r}\frac{d}{dr}\right)A_{0}(r)  
+e\rho_{\rm v,p}(r)=0$, and it only depends explicitly on the photon field $A_0(r)$ and on the protons components $g(r)$ and $f(r)$ through the density $\rho_{\rm v,p}(r)$. There is a different associated operator for each proton and neutron level for the Dirac equations, but we omit a tracking index to keep the notation simpler.

Finding a solution $\Xi$ for given parameters $\alpha$ means finding a collection of fields and wave functions such that all the operators $F^{(\cdot)}_\alpha[\Xi]$ acting on such list give back the function that is zero for every $r$. Such solution must satisfy as well the normalization condition \eqref{Eq: norm}. In general, these requirements cannot be satisfied by any choice of the RB coefficients under the RB approximation, i.e. $F^{(\cdot)}_\alpha[\hat \Xi]\neq 0$ simultaneously for any choice of $\boldsymbol{a}$. We can relax these conditions by projecting each operator $F_\alpha[\hat \Xi]$ over a set of ``judges" $\psi_j^{(\cdot)} (r)$ \cite{Bonilla:2022rph} and requiring that the projections are zero:

\begin{subequations}
\begin{eqnarray}
 &&
\langle \psi_j^\Phi  |  F_\alpha^\Phi[\hat\Xi]\rangle=0, \quad 1\leq j \leq n_\Phi,  \\
 &&
 \langle \psi_j^W |F_\alpha^W[\hat\Xi]\rangle=0,  \quad 1\leq j \leq n_W,  \\
 &&
\langle \psi_j^B  |F_\alpha^B[\hat\Xi]\rangle=0, \quad 1\leq j \leq n_B,  \\
 &&
\langle \psi_j^A  |F_\alpha^A[\hat\Xi]\rangle=0,  \quad 1\leq j \leq n_A, \label{Eq: projections A}\\
 &&
\langle \psi_j^g |F_\alpha^g[\hat\Xi]\rangle=0, \quad 1\leq j \leq n_g, \label{Eq: projections G} \\
 &&
\langle \psi_j^f |F_\alpha^f[\hat\Xi]\rangle=0, \quad 1\leq j \leq n_f \label{Eq: projections F},
\end{eqnarray}
\label{Eq: projections}
\end{subequations}

in which we have made the choice of projecting each operator $F^{(\cdot)}_\alpha[\Xi]$ a total of $n_{(\cdot)}$ times, where  $n_{(\cdot)}$ is the number of RB expanding the associated function. Once again, there will be a different set of projection equations for every proton and neutron level for a given nucleus. The projection operation, which we write using Dirac's notation, means the regular functional integral used in Hilbert spaces, for example:
\begin{equation}
    \langle \psi(r) | \phi(r) \rangle \equiv \int \psi^{*}(r)\phi(r)dr
\end{equation}

Following our previous approach \cite{Bonilla:2022rph}, we choose the ``judges'' to be the same as the RB expansion, as it is common practice with the Galerkin method~\cite{fletcher1984computational}. For example, in the case of $^{48}$Ca with three basis for every field and wave function, since the photon field RB expansion $\hat A_0 = \sum_{k=1}^3 a_k^A A_k$ has three unknown coefficients, there will be three ``judges" projecting the operator $F_\alpha^A[\hat\Xi]$. These ``judges" are chosen as $\psi_j^A(r)=A_j(r)$ for $1\leq j\leq 3$. In total, for this $^{48}$Ca example, we will have three projection equations for each field, three equations for each $g$ and three for each $f$ for each level of protons and neutrons, for a total of 90 projection equations. Such system of equations, together with the normalization conditions uniquely determines (if it has a solution) the 103 RB coefficients and energies $\boldsymbol{a}$ for each new value of the parameters $\alpha$.

We also note that the dependence of the equations $\eqref{KGEqns}$ and $\eqref{DiracEqn}$ on the parameters $\alpha$ is affine, which means that every operator  can be separated into a product of a function that only depends on $\alpha$ and a function that only depends on $r$. For example, the non-linear coupling between the isoscalar-vector meson $\omega$ and the isovector-vector meson $\rho$ in Eq.~\eqref{KGEqns: B} reads: $-2\big(g_\rho^2\Lambda_\text{v} \big) \big[W_0(r)B_0(r) \big]$. In practice, this means that every integral in $r$ in the projection equations~\eqref{Eq: projections} can be done, once the RB has been fully specified, without explicitly assigning numerical values to the parameters such as $g_\rho^2$ or $\Lambda_\text{v}$. These computations are usually done once during the offline stage and then stored in memory to be used during the online stage \cite{hesthaven2016certified}. The result of this procedure is a set of projection equations---agnostic to the $r$ variable---that do not involve any computation in the high fidelity space of size $\mathcal{N}$. These equations will involve a small amount of linear combinations of products of the model parameters $\alpha$ and the unknown coefficients $\boldsymbol{a}$, usually much more computationally tractable than the original coupled equations of size $\mathcal{N}$. The two observables we study in this work, the charge radius and binding energy of each nucleus, are also affine functions of the parameters $\alpha$ and the solution's coefficients $\boldsymbol{a}$, see Eqs.~\eqref{Eq: Radius},~\eqref{Eq: Nuc En},~\eqref{Eq: En Fields}, and~\eqref{Eq: Field Energy}. This means that these observables can also be pre-computed, avoiding calculations of size $\mathcal N$ when the emulator is used for fast evaluations\footnote{If the dependence on the parameters $\alpha$ of the operators involved in the system's equations, or in the observable's calculations is not affine, techniques such as the Empirical Interpolation Method \cite{hesthaven2016certified,grepl2007efficient,barrault2004empirical} can be implemented to avoid computations of size $\mathcal{N}$ in the online stage.}. 

For a concrete example, consider $^{48}$Ca now with only 2 basis for all functions. Each of the two projection equations associated with the proton's $g$ and $f$ \eqref{Eq: projections G} and \eqref{Eq: projections F} contains a total of 22 terms. The equation for $f$ for the first proton level---with the choice of basis we describe in the next section---with all numbers printed to two decimals precision reads:
\begin{equation}
\begin{split}
    &\langle g_1^f(r) |F_\alpha^f[\hat\Xi]\rangle = \\
    &0.06a^f_1-0.09a^f_2-E^P(0.8a^g_2 +1.66a^g_1) + \\
    & M(1.66a^g_1 +0.8a^g_2)-2.89a^\Phi_1a^g_1-1.45a^\Phi_2a^g_1+\\
    &2.35a^W_1a^g_1+1.17a^W_2a^g_1- 0.03a^B_1a^g_1-0.01a^B_2a^g_1+ \\
    &0.07a^A_1a^g_1-0.01a^A_2a^g_1-0.8a^\Phi_1a^g_2-2.71a^\Phi_2a^g_2+\\
    &0.64a^W_1a^g_2+2.18a^W_2a^g_2-  0.02a^B_1a^g_2+0.01a^B_2a^g_2+\\
    &0.03a^A_1a^g_2+0.04a^A_2a^g_2 =0.
\end{split}
\label{Eq: poly}
\end{equation}

\subsection{Accuracy vs Speed: Basis selection}

As with many other computational methods, the RBM posses a trade-off between accuracy and speed. If we use more bases for the expansions~\eqref{Eq: RB} we expect that our approximation will be closer to the high fidelity calculation, but that will come at the expense of more coefficients to solve for in the projection equations~\eqref{Eq: projections}. If we use too few bases, the underlying physical model will be miss-represented when compared with experimental data, but if we use too many bases we might waste computational time to obtain an unnecessary accuracy level. To find a satisfactory balance we study the performance of the RBM, both in terms of accuracy and speed, for different basis size configurations on a validation set containing 50 new high fidelity evaluations drawn from the same distribution as the one we used for the training set \cite{Chen:2014sca}.

As a metric of performance we define the emulator root mean squared errors as:
\begin{equation}\label{Eq: RMSE R}
    \Delta \text{R}_\text{ch} = \sqrt{\frac{1}{N_v}\sum_{i=1}^{N_v}\Big(\text{R}^\text{mo}_\text{ch\ i}-\text{R}^\text{em}_\text{ch \ i}\Big)^2}
\end{equation}

for the charge radius, and as:

\begin{equation}\label{Eq: RMSE BE}
    \Delta \text{BE} = \sqrt{\frac{1}{N_v}\sum_{i=1}^{N_v}\Big(\text{BE}_i^\text{mo}-\text{BE}^\text{em}_i\Big)^2},
\end{equation}

for the total binding energy. In both expressions  $N_v$ is the total number of samples in the validation set (50) and the superscripts ``mo" and ``em" stand for the high fidelity model and the RBM emulator, respectively.

A straightforward approach for exploring different basis configurations consists of setting all basis numbers $\{n_\Phi,n_W,n_B,n_A,n_g,n_f\}$ to the same value $n$.
There are two main disadvantages of this approach. First, the basis increments can be too big, making it harder to obtain a good trade-off. For example, in the case of $^{208}$Pb with $n=2$ we have a total of 160 basis, while for $n=3$ we jump straight to 240. Second, the accuracy in the emulation of the observables could be impacted differently by how well we reproduce each function involved in~\eqref{KGEqns} and~\eqref{DiracEqn}. Having a leverage that allows us to dedicate more resources (bases, that is) to more crucial functions could be therefore beneficial, and the simplistic approach with a common number $n$ is unable to optimize the computational resources in that sense.

On the other hand, exploring all possible basis size configurations for a given maximum basis size is a combinatorial problem that can quickly become intractable. Therefore, we decided to follow a Greedy-type optimization procedure in which we incrementally add new basis to the current configuration, choosing the ``best" local option at each step. The basis are chosen from the principal components obtained from the training set of 50 high fidelity runs. For all the nuclei the starting configuration was 7 basis for each one of the fields $\{ \Phi_0, W_0, B_0, A_0 \}$ and 2 basis for each of the wave functions $g$ and $f$ on all the nucleus' levels. On each step, we add one more basis to both $g$ and $f$ to the four levels across both protons and neutrons which were reproduced most poorly in the previous iteration on the validation set. The ``worst performers" are chosen alternating in terms of either the single particle energies (serving as a proxy for the total binding energy), and the $L^2$ norm on the wavefunctions themselves (serving as a proxy for the proton and neutron radius). The fields basis numbers are all increased by 1 once one of the wave functions basis number reaches their current level (7 in this case).

\begin{figure}[h]
    \centering
    \includegraphics[width=0.48\textwidth]{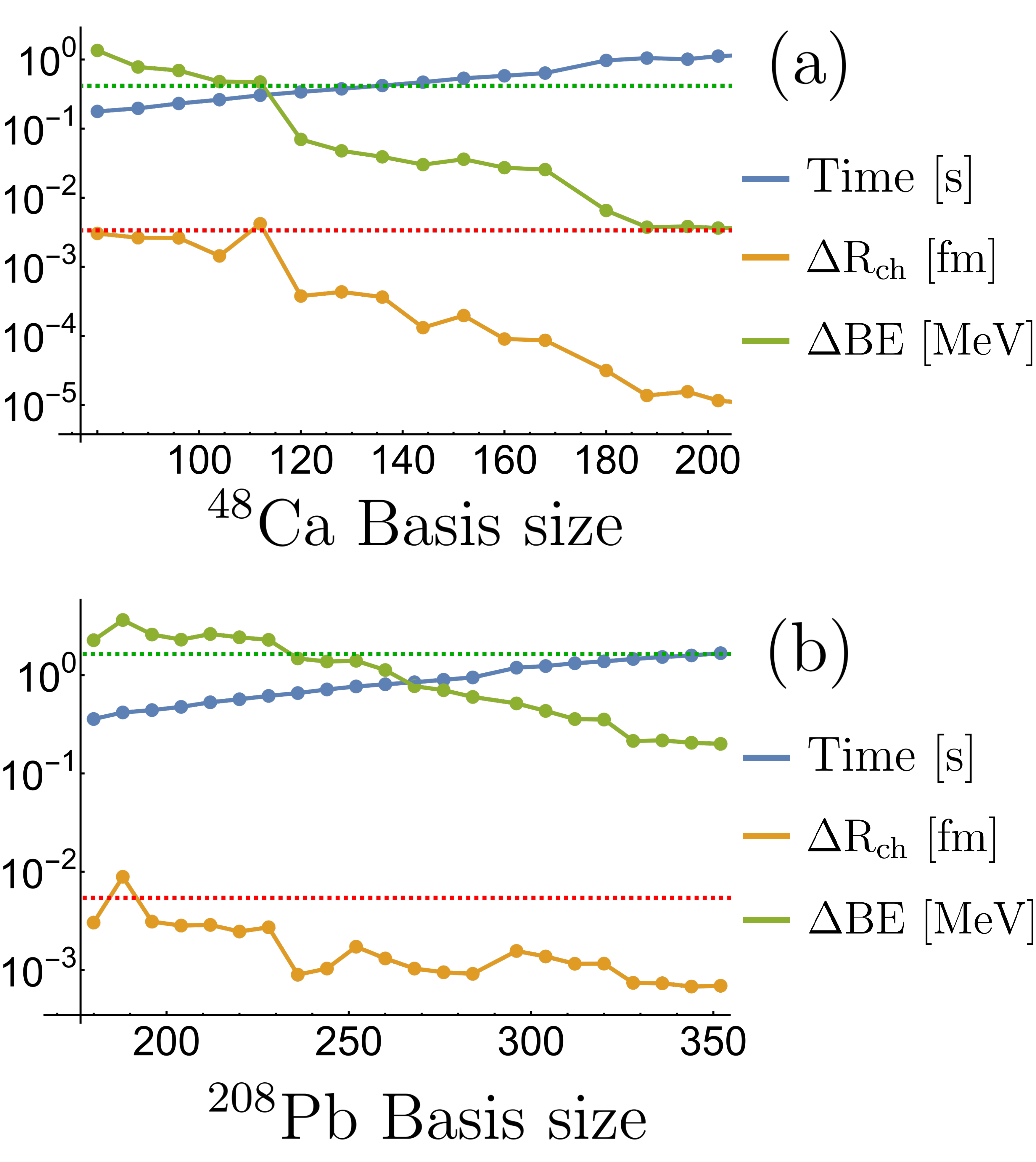}
    \caption{Performance of the RBM emulator for $^{48}$Ca (a) and $^{208}$Pb (b) as the total number of basis is increased following the Greedy algorithm described in the text. The dashed red and green lines in both plot indicate an error of $0.1\%$ in the charge radius and total binding energy, respectively. The computation time per sample is calculated solving the RBM equations in Mathematica, which is substantially slower than the production emulator used in the calibration and detailed in Sec.~\ref{Subsec: Code}.}
        \label{fig:Basis Training}
\end{figure}

For example, for $^{48}$Ca we start with 7 basis for the four fields $\{n_\Phi,n_W,n_B,n_A\}=\{7,7,7,7\}$, and $\{n_g,n_f\}=\{2,2\}$ for every one of the six levels for protons and seven levels for neutrons. On the first step we compare the RBM calculations with the 50 high fidelity solutions from the validation set and identify the first neutron level, and the first, third, and fifth proton levels as the worse (on average) estimated single particle energies. Consequently, their respective basis are increased by their third principal components (see Figure~\ref{fig:CaPCA}): $\{n_g,n_f\}=\{3,3\}$. On the next step, we re-calculate the RBM solutions with the new updated basis and identify the fifth neutron level, and the second, third, and sixth proton levels as the worst performances in terms of the overall wave function sense (the $L^2$ norm). This procedure is repeated as the overall basis number increases as is illustrated in Figure~\ref{fig:Basis Training} for $^{48}$Ca and $^{208}$Pb. We observed similar behaviors for the other 8 nuclei involved in this study.

For the range of basis explored, the overall performance for both the charge radius and total binding energy roughly improves exponentially, although not monotonically\footnote{For some steps the emulator's performance -in terms of $\Delta$ R$_\text{ch}$ and $\Delta$ BE- gets worse when adding the 4 new basis, which at first might seem counter-intuitive. It is important to note, however, that with each new basis we add we are changing the entire system of equations both by adding four new projections and by adding new elements to the previous ones. Nothing prevents the solution $\boldsymbol{a}$ to the new system to under perform in comparison to the previous one in the particular metric we are using.}, with the total basis number. The error in the radius and binding energy reduces by more than a factor of 100 for $^{48}$Ca and by a factor of 10 for $^{208}$Pb. The computational time also increases exponentially, expanding almost an entire order of magnitude for both nuclei.

To select an optimal configuration of bases we used as a target a $0.1\%$ error in both observables for all nuclei involved, which is roughly three times smaller than the average deviation between the originally calibrated RMF and the available experimental values \cite{Chen:2014sca}. These targets are shown as the red and green dashed lines in Figure~\ref{fig:Basis Training}.
Table~\ref{Table: Emulator Results} shows the basis size for the chosen basis and the results from this validation analysis. In the case where we would like to have a faster emulator at the expense of accuracy, we could choose a smaller basis size from the configurations showed in Figure~\ref{fig:Basis Training}. In the case where we need a more accurate emulator for particular calculations, configurations with more basis functions could be chosen at the expense of speed.

\begin{table}[]
\begin{tabular}{ccccccc}
Nuc. & \begin{tabular}[c]{@{}c@{}}Basis\\ Size\end{tabular} & \begin{tabular}[c]{@{}c@{}}Time\\ {[}ms{]}\end{tabular} & \begin{tabular}[c]{@{}c@{}}$\Delta$ R$_\text{ch}$\\ {[}$10^{-3}$ fm{]}\end{tabular} & \begin{tabular}[c]{@{}c@{}} $0.1\%$ R$_\text{ch}$\\ {[}$10^{-3}$ fm{]}\end{tabular}  & \begin{tabular}[c]{@{}c@{}}$\Delta$ BE\\ {[}MeV{]}\end{tabular} &  \begin{tabular}[c]{@{}c@{}}$0.1\%$ BE\\ {[}MeV{]}\end{tabular} \\ \hline
$^{16}$O   & 68          & 0.7            & 1.8                & 2.7                  & 0.1     & 0.1         \\
$^{40}$Ca  & 116         & 2.2            & 1.2               & 3.5                  & 0.2     & 0.3         \\
$^{48}$Ca  & 120         & 2.4            & 0.4               & 3.5                  & 0.1     & 0.4         \\
$^{68}$Ni  & 128         & 3.1            & 1.8               & 3.9                  & 0.5     & 0.6         \\
$^{90}$Zr  & 168         & 6.6            & 0.9               & 4.3                  & 0.2     & 0.8         \\
$^{100}$Sn & 180         & 8.1           & 1.0               & 4.5                  & 0.3     & 0.8         \\
$^{116}$Sn & 176         & 8.0            & 2.4               & 4.6                  & 0.8     & 1.0         \\
$^{132}$Sn & 184         & 9.5            & 1.9               & 4.7                 & 0.8     & 1.1          \\
$^{144}$Sm & 216         & 14           & 1.8               & 4.9                  & 0.8     & 1.2          \\
$^{208}$Pb & 236         & 20            & 0.9                & 5.5                  & 1.5     & 1.6        
\end{tabular}

\caption{Results from the basis selection procedure using the 50 samples from the validation set. The second column shows the total basis size for the selected configuration for each nucleus for the RBM emulator we use in the rest of the manuscript. Column three shows the average time to compute a single RB full solution for that nucleus using the optimized compiled emulator in Python, which we detail in the next section. Columns four and six show the root mean squared error of the emulator (see Eqs.~\eqref{Eq: RMSE R} and~\eqref{Eq: RMSE BE}) when compared to the high fidelity solutions for the charge radius and the total binding energy, respectively. Columns five and seven show the target of $0.1\%$ of the experimental value of the respective quantity used in the basis selection procedure. For the charge radius of $^{68}$Ni and $^{100}$Sn the central value of FSUGold2 \cite{Chen:2014sca} was used instead for column five.} \label{Table: Emulator Results}
\end{table}

\subsection{RBM code optimization}\label{Subsec: Code}

The offline stage consisting of the symbolic construction of the Galerkin projection equations and the expressions for the observables of interest is performed in Mathematica, resulting in polynomial equations in the parameters and RB coefficients such as Eq.~\eqref{Eq: poly}.
These equations are then parsed and converted into both Python and Fortran functions which can then be compiled into a library and evaluated in the calibration driver software.
The explicit Jacobian matrix is also constructed and parsed in the same way, resulting in fewer evaluations of the polynomial equations and thus faster convergence of the root finding routine.
This automated pipeline from symbolic representation in Mathematica to compiled Python library vastly simplifies the development process of the emulator and allows for various basis sizes to be included at will while ensuring an efficient implementation.

For the Python implementation Cython is used to first convert the Python code into C which is subsequently compiled into a Python compatible library.
The Fortran implementation is similarly compiled using the \texttt{NumPy} \texttt{f2py} tool to produce a performant Fortran library with an appropriate Python interface.
Regardless of the generating code, the resulting interface of the modules are the same and can be used interchangeably depending on the needs of the user.
Each evaluation of the emulator for a given set of parameters then uses the MINPACK-derived root finding routine in \texttt{SciPy}~\cite{scipy} to find the optimal basis coefficients which are used as input for the observable calculations.

This procedure results in a time-to-solution on the order of hundreds of microseconds to tens of milliseconds depending on the nucleus being considered, as detailed in Table~\ref{Table: Emulator Results}.
The Runge-Kutta high-fidelity solver (written in Fortran) does not exhibit such a strong scaling across different nuclei, thus the relative speed-ups vary
from 25,000x for $^{16}$O, to 9,000x for $^{48}$Ca and 1,500x for $^{208}$Pb.
This level of performance brings the evaluation of the surrogate model well within our time budget for the calibration procedure and also represents the simplest method in terms of developmental complexity.
If the evaluation of the emulator needs to be further accelerated, a pure Fortran implementation of the root finding routine exhibits an additional decrease in time-to-solution of order 3x in comparison to the hybrid Python/compiled model detailed above at the cost of a slightly less user-friendly interface for the emulator.

Having constructed an emulator with the accuracy and calculation speed level we require, we now proceed to build the Bayesian statistical framework that will be used to perform the model calibration. In this calibration, the emulator finite accuracy will be included as part of our statistical model.

\section{Framework for Bayesian Uncertainty Quantification} \label{Sec: Bayesian}

To calibrate our nuclear model properly, we need to account for the sources of error associated with each data point. We will use the well-principled Bayesian framework for this task \cite{gelman1995bayesian,phillips2021get}, which produces a full evaluation of uncertainty for every model parameter, in the form of posterior probability distributions given all the data. Its ingredients are twofold: first, a probability model, known as the likelihood model, for the statistical errors linking the physically modeled (or emulated) output to the experimental data given physical model parameters; second, another probability model for one's a-priori assumptions about the physical model parameters.
The output of the Bayesian analysis takes the form of a probability distribution for all model parameters; this is known as the parameters' posterior distribution. In this work, given the paucity of data, we choose to estimate the standard deviations of the statistical models separately, ahead of the Bayesian analysis, either using uncertainty levels reported in the literature, or using a natural frequentist statistical estimator. This minor deviation from a fully Bayesian framework is computationally very advantageous, an important consideration given this manuscript's overall objective. 

The Bayesian framework can also be used as a predictive tool, by integrating the high-fidelity or emulated physical model against the posterior distribution of all model parameters, for any hypothetical experimental conditions which have not yet been the subject of an experimental campaign. Such predictions are expected also to take into account the uncertainty coming from the statistical errors in the likelihood version of the physical model. Relatively early examples of these uses of Bayesian features in nuclear physics work can be found in \cite{Neufcourt:2018syo, Neufcourt:2019qvd}.

In this section, we provide the details of our likelihood and prior models, and how they are built in a natural way, as they relate to experimental values, their associated emulated values, and all physical model parameters. We also explain in detail how the statistical model variance parameters are estimated ahead of the Bayesian analysis. All our modeling choices are justified using the physical context and the simple principle of keeping statistical models as parsimonious as possible.

\vspace{.2in}

\subsection{Specification of the statistical errors and the likelihood model}\label{subsec: Like}

\begin{figure}[h]
    \centering
    \includegraphics[width=0.48\textwidth]{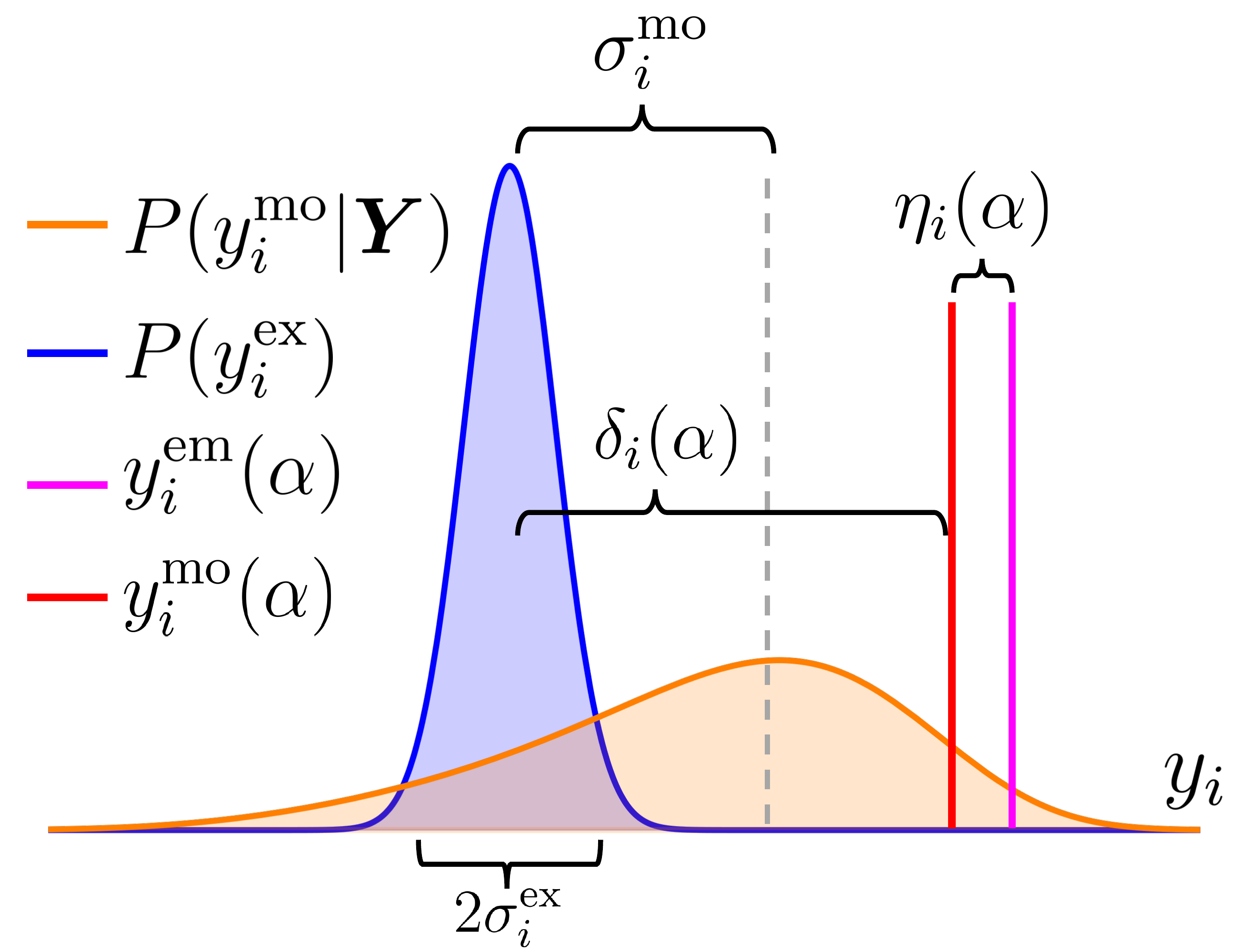}
    \caption{Visual representation of the statistical model with the three sources of uncertainty for an observable $y_i$. For a particular value of $\alpha$, the model calculation, $y_i^\text{mo}$, (red vertical line) deviates from the center of the experimental distribution, $P(y_i^\text{ex})$, (blue curve) by the model error $\delta_i(\alpha)$. The size of the experimental error, characterized by $\sigma_i^\text{ex}$, is exaggerated in the figure to facilitate showing. The estimated value by the emulator, $y_i^\text{em}$, (vertical magenta line) deviates from the model calculation, $y_i^\text{mo}$, by the emulator error, $\eta_i(\alpha)$. The model error scale, $\sigma_i^\text{mo}$, characterizes the expected size of $\delta_i(\alpha)$ as the parameters $\alpha$ are varied within their meaningful physical range. This parameter range is characterized by the Bayesian posterior distribution $P(y_i^\text{mo}|\boldsymbol{Y})$ (orange curve), -obtained only after the analysis is done- of the observable $y_i$ given all the calibration data $\boldsymbol{Y}$ .}
        \label{fig:Stat errors}
\end{figure}

Let us denote by $y^\text{ex}_i$ the $i$-th experimental observation --- binding energies or charge radii in our case --- of the 10 nuclei considered. We have a total of 10 measured binding energies and 8 charge radii ($^{68}$Ni and $^{100}$Sn do not have measured charge radii), therefore $1\leq i\leq18$. Let us denote by $y_i^\text{mo}(\alpha)$ the high fidelity model calculation associated with $y^\text{ex}_i$ for a given value of the model parameters $\alpha$. Finally, let us denote $y_i^\text{em}(\alpha)$ the RBM emulated calculation associated with the same observable. We identify three main sources of errors\footnote{A fourth source of error could be, in principle, the computational error in the high fidelity solver for the physical model. We expect this error to be negligible in comparison to the other three at the level of resolution $\Delta r$ our high fidelity solvers have.} in the model calibration, namely experimental, modeling,
and emulation errors --- the latter being the difference between $y^\text{mo}_i(\alpha)$ and $y^\text{em}_i(\alpha)$ --- which we write into a statistical model as follows: for every $i$,
\begin{equation}\label{likelihood model}
\begin{split}
y^\text{ex}_i&= y_i^\text{mo}(\alpha) + \delta_i(\alpha) + \epsilon_i \\
& =y_i^\text{em}(\alpha) + \eta_i(\alpha) +\delta_i(\alpha) + \epsilon_i.
\end{split}
\end{equation}
These three sources of error are represented in Figure~\ref{fig:Stat errors} as an illustrative stylized example.

The experimental error, $\epsilon_i$, is assumed to  come from a normal distribution with mean zero and standard deviation $\sigma^\text{ex}_i$: $\epsilon_i\sim \mathcal{N}(0,\ (\sigma^\text{ex}_i) ^2)$. These errors are assumed to be uncorrelated between measurements of different nuclei and different quantities. The error scale for each measurement, $\sigma^\text{ex}_i$, is an estimate of the aggregate of the many uncertainty sources --- both systematic and statistical --- that can play a role during the experimental campaign. In principle, since each measurement comes from a different campaign $i$, it is important to allow $\sigma^\text{ex}_i$ to change from $i$ to $i$. Since the experiments are not conducted in consort, it is legitimate to assume these errors are uncorrelated. As shown in Figure~\ref{fig:Stat errors}, these experimental errors $\epsilon_i$ should not be interpreted as the discrepancy between the theoretical prediction and the experimental value. Rather, they represent the estimated difference between the observed experimental value during realistic conditions, compared to its (unattainable) value in ideal settings free of experimental noise. This noise is thus due for instance to the known imprecision of measurement instruments.

The modeling error, or model discrepancy term, $\delta_i(\alpha)$, represents the intrinsic failures of the physical model when reproducing reality, for a given value of $\alpha$. It is an aggregate of the many simplifications made during the construction of the model, as well as any physical effects, known or unknown, which are unaccounted for by the physical model. In the limit where the experimental errors become negligible, it is the term that explains the deviation between theory and observation. Due to these model limitations, we expect that even the best regions in the parameter space --- values of $\alpha$ that make the discrepancies as small as possible --- cannot make them all vanish simultaneously ($\delta_i(\alpha)=0$) for every observable $i$.

It is typically unrealistic to expect a very precise estimate of the statistical properties of the set of $\delta_i(\alpha)$ as $\alpha$ varies. Indeed, first, because the usual dataset in low-energy physics studies consist only of a few spherical magic or semi-magic nuclei, limiting the statistical analyses that can be made. Second, since the origin of $\delta_i(\alpha)$ roots in phenomena which we do not dominate completely, it becomes very hard to give accurate estimates when the experimental observations are not available. This motivates us to propose a parsimonious model, in accordance with the statistical principle that parsimony promotes robustness, an idea that traces back several decades (e.g. \cite{Ludwig:1983}).  
We assume that, up to scaling at the level of observables, the modeling error variances are shared within each of the two observable categories (binding energies and charge radii), and do not depend on the parameters $\alpha$ within their physical meaningful range that reproduces the nuclear properties. This is represented by the scale $\sigma_i^\text{mo}$ (to be defined precisely shortly) in Figure~\ref{fig:Stat errors}.

Finally, for a given fixed value of the parameters $\alpha$, the emulator error, $\eta_i(\alpha)$, represents the difference between the model's original high fidelity calculation, and the approximate version computed by the emulator. In Figure~\ref{fig:Stat errors} it is represented as the difference between the red and magenta vertical lines. This is the easiest error to obtain exactly, given a fixed $\alpha$, since it is entirely computable, given access to the high fidelity and the emulator implementations. The challenge lies in estimating $\eta_i(\alpha)$ for new values of $\alpha$ \emph{without} the use of the high fidelity solver. In the RBM literature, there exist approaches to estimate the emulator's error in terms of the properties of the underlying differential equation \cite{hesthaven2016certified,veroy2003posteriori,buffa2012priori}, but to our knowledge they have not been yet extended to the type of coupled nonlinear equations that describe our physical model~\eqref{KGEqns} and~\eqref{DiracEqn}. Our proposal below is to model all emulator errors, including the unobserved ones, using the same statistical model, where the emulator error intensity does not depend on $\alpha$, thereby circumventing the issue of developing an analytical approach to extrapolating these errors in a non-linear setting, and keeping with the principle of parsimony.  

Having identified and described these three sources of errors, we proceed to propose and implement methods to estimate their combined effect in order to  calibrate our physical model properly through the RBM emulator. 

In the case of binding energies and charge radii, the experimental determinations are precise enough that the typical error scale $\sigma^\text{ex}_i$ can be ignored in comparison to the typical model discrepancies\footnote{For example, the binding energy of $^{208}$Pb is known to a precision better than $10^{-4}\%$ \cite{huang2021ame}, while its charge radius to a precision of 0.02$\%$ \cite{angeli2013table}. In contrast, the estimated model error we calculate in the following discussion for the same quantities is
$0.25\%$
and $0.26\%$, respectively.}. 
Therefore, we decide to neglect the experimental errors for the rest of our analysis. 

We assume that the model discrepancies $\delta_i(\alpha)$ scale proportionally to the value of each individual experimental datapoint. This is because each datapoint represents a different physical reality, and while, say, two binding energies for two similar nuclei may be subject to the same intensity of modeling error, this may not be a good assumption for two nuclei which are more distant in the nuclear landscape.
Specifically, we assume that each scaled model discrepancies $\tilde \delta_i \equiv \delta_i/y_i^\text{ex}$  comes from a common normal distribution with mean zero and either variance $\sigma^2_\text{BE}$ for a binding energy datapoint or variance $\sigma^2_\text{R}$ for a charge radius datapopint. We also assume that these errors are independent of each other, and thus uncorrelated. We estimate these errors' scales $\sigma_\text{BE}$ and $\sigma_\text{R}$ from the deviations between the originally calibrated RMF model FSUGold2~\cite{Chen:2014sca} and the experimental observations, simply by using a version of the classical unbiased variance estimator.  Explicitly, for the variance of the modeling errors on the binding energy side,  where the model is FSUGold2, with $N_\text{BE} = 10$ for the ten binding energy datapoints, we let 
\begin{equation}\label{MVUE}
\sigma_\text{BE}^2 = \frac{1}{N_\text{BE}}\sum_{i=1}^{N_\text{BE}} \Big(\frac{y_i^\text{ex}-y_i^\text{FSUGold2}}{y_i^\text{ex}}\Big)^2,
\end{equation}
and similarly for $\sigma_\text{R}^2$ as the variance of the modeling error for charge radii. These expressions are calculated for all the data with available experimental values in Table II in \cite{Chen:2014sca}. These formulas are the classical minimum-variance unbiased estimators (MVUE) of variances for datapoints coming from a normal distribution with known means and unknown common variance. 
One can view each model-calculated datapoint as the error-prone data, with the experimental value as its mean value. This results in a mathematically identical unbiased estimator as if all means were equal. In our case, since we choose to normalize the modeled data by dividing it by the experimental data, we are in fact handling a classical situation, where the data's mean value is known to equal 1. In that scenario, the classical MVUE is the one given in formula \eqref{MVUE}. Note that its leading factor is $1/N_\text{BE}$ rather than $1/(N_\text{BE}-1)$; this is because the mean is known. In other words, the model to which this MVUE \eqref{MVUE} responds is
\begin{equation}
\frac{y_i^\text{FSUGold2}}{y_i^\text{ex}}=1+\tilde \delta_i
\end{equation}
where $\tilde \delta_i$ are assumed to be independent mean-zero normal errors with unknown variance $\sigma^2_\text{BE}$ for the binding-energy data, and similarly for $\sigma^2_\text{R}$. Applying the estimation to the data in \cite{Chen:2014sca} we obtain:
\begin{equation}
\sigma_\text{BE}=0.25\%,
\end{equation} 
and 
\begin{equation}
\sigma_\text{R}=0.26\%.
\end{equation} 

We express these two values as percentages since they are dimensionless. We treat the charge radius of $^{16}$O as an outlier and exclude it from this estimation, assigning it its own estimated error scale of $\sigma_{\text{R},^{16} \text{O}}=1.4\%$, so that $N_\text{R}=7$. The corresponding modeling error standard deviations $\sigma_{i}^\text{mo}$ for specific observables $y_i$ are obtained by multiplying the one based on scaled data by their respective experimental values, which provides the correct standard deviations for all $\delta_i$ in accordance with how we defined $\tilde \delta_i$. For example, for $i=$ BE for $^{48}$Ca, $\sigma_i^\text{mo}=\sigma_\text{BE} \times \big(416\ \text{MeV}\big)=\frac{0.25}{100}\times \big(416\ \text{MeV}\big)\approx 1$ MeV.\footnote{The experimental values of the charge radii for $^{68}$Ni and $^{100}$Sn are not known. In these cases, we used the values reported for FSUGold2, as proxies, to preserve these two nuclei in the analysis when creating predictive posterior distributions with the calibrated model.} 

We follow a similar parsimonious approach and model the emulator error $\eta_i(\alpha)$ as coming from a normal distribution with mean zero and scale (standard deviation) $\sigma_i^\text{em}$ that does not depend on $\alpha$. In Figure~\ref{fig:Stat errors}, $\sigma_i^\text{em}$ would be the scale of a Gaussian distribution (not shown to keep the figure easier to read) centered at $y_i^\text{em}(\alpha)$. From our assumptions we would expect that such distribution will contain the true model evaluation $y_i^\text{mo}$ around $68\%$ of the time both computations are made, \emph{independent} of the exact value of $\alpha$ within the physically meaningful range where the emulator was trained.

We estimate the scale $\sigma_i^\text{em}$ from the empirically observed deviations between the RBM emulator and the high fidelity solutions in the validation set used for the selection of the basis in the previous section. We select, therefore, $\sigma_i^\text{em}$ as the values reported in the fourth column ($\Delta$ R$_\text{ch}$) and sixth column ($\Delta$ BE) in Table~\ref{Table: Emulator Results}. Given the extraordinary performance that the RBM and similar emulator approaches have shown when extrapolation is involved \cite{Bonilla:2022rph,frame2018eigenvector,konig2020eigenvector}, we expect the assumption that $\sigma_i^\text{em}$ does not depend on $\alpha$ to be valid unless we evaluate the RBM emulator too far away from its training region. These regions are highly unlikely to be visited by the Monte Carlo sampling we use later for the model calibration. Finally, we assume that the emulator's errors and the model discrepancy errors are independent (and thus uncorrelated) across different quantities, and within the same observable as well.

Under the assumptions we have made about the three sources of uncertainties, including the independence of all error terms which implies that the variance of their sum is the sum of their variances, we can finally specify the likelihood function for our statistical model. To simplify the prior specification and the exploration of the parameter space, we construct our statistical modeling using the bulk matter parametrization $\theta$, which is equivalent to the Lagrangian couplings one with $\alpha$ (see Sec.~\ref{Sec:RMF}). Denoting $N=N_\text{BE}+N_\text{R} = 18$ and denoting by $\boldsymbol{Y}$ the $N$-dimensional
vector formed of the experimental datapoints $y_i^\textbf{ex}$,
 our likelihood model is: 
\begin{equation}\label{likelihood}
    P(\boldsymbol{Y}|\theta) \propto e^{-\chi^2/2},
\end{equation}
where
\begin{equation}
    \chi^2 = \sum_{i_1=1}^{N_\text{BE}} \frac{(y^\text{em}_{i_1}(\theta)-y^\text{ex}_{i_1})^2}{\sigma_{i_1}^\text{em}\ ^2 + \sigma_{i_1}^\text{mo}\ ^2} +    \sum_{i_2=1}^{N_\text{R}} \frac{(y^\text{em}_{i_2}(\theta)-y^\text{ex}_{i_2})^2}{\sigma_{i_2}^\text{em}\ ^2 + \sigma_{i_2}^\text{mo}\ ^2}.
\end{equation}
Our modeling assumptions about the error structure, plus the standardization in $\boldsymbol{Y}$, imply indeed that $\chi^2$ is chi-squared distributed with $N$ degrees of freedom. Note that $\sigma_{i_2}^\text{mo}$ has the associated $0.26\%$ value for all $i_2$ except for $^{16}$O, for which it has the value of $1.4\%$.

\subsection{Prior}

For the prior distribution we adopt an uncorrelated multivariate normal in $\theta$ as follows: 
\begin{equation}\label{prior}
    P(\theta) \propto e^{-\chi^2_0/2},
\end{equation}
where:
\begin{equation}
    \chi^2_0 = \sum_{j=1}^8 \frac{(\theta_j-\theta_{0,j})^2}{\sigma_{\theta, j}^2}.
\end{equation}
The central values $\theta_{0,j}$ and standard deviations $\sigma_{\theta,j}$ are specified for the eight components of $\theta$ in Table~\ref{Tab: Prior values}. They were chosen to roughly cover the expected parameter region with wide ranges based on the previous calibration \cite{Chen:2014sca} .

\begin{table}[]
\begin{tabular}{ccc}
$\theta_j$                             & $\theta_{0,j}$ & $\sigma_{\theta,j}$ \\ \hline
$M_s$ {[}MeV{]}                      & 500        & 50              \\
$\rho_0$ {[}fm$^{-3}${]}                & 0.15        & 0.04             \\
$\epsilon_0$ {[}MeV{]}               & -16        & 1             \\
$M^{*}$ {[}MeV{]}                    & 0.6        & 0.1             \\
$K$ {[}MeV{]}                        & 230        & 10              \\
$\zeta$ \ \ \ \  \ \ \ \ \ \  & 0.03       & 0.03            \\
$J$ {[}MeV{]}                        & 34         & 4               \\
$L$ {[}MeV{]}                        & 80         & 40             
\end{tabular}\caption{Prior central values and standard deviations for the eight model parameters used in the calibration. }\label{Tab: Prior values}
\end{table}

\subsection{Posterior}

With our likelihood and priors fully set up, the posterior densities
$P(\theta | \boldsymbol{Y})$ for the parameters $\theta$ are given classically by Bayes' rule \cite{gelman1995bayesian} as being proportional to the product of the likelihood in \eqref{likelihood} and the prior in \eqref{prior}, where the likelihood is evaluated at the experimentally observed datapoints labeled above as $\boldsymbol{Y}$. From here, we are interested in using the Bayesian analysis to
compare the calculations of the fully calibrated model with the experimental values of the observables, to verify that our uncertainty quantification is accurate. If our uncertainty bands on these predicted values are too narrow (too optimistic), too high a proportion of our 18 observations will fall outside of the bands. If our uncertainty bands are too wide (too conservative), many or all of our 18 observations will be inside their corresponding uncertainty bands. Being slightly too conservative is easily construed as a virtue to hedge against the risk of being too optimistic. The latter should be construed as an ill-reported uncertainty quantification. The method we propose here, to gauge the accuracy our uncertainty quantification, with results described in Section 5, is a manual/visual implementation of the now classical notion of Empirical Coverage Probability (ECP, see \cite{Neufcourt:2018syo} for a nuclear physics implementation), appropriate for our very small dataset with 18 points. To compute the posterior density of every predicted value corresponding to our experimental observations, we view the likelihood model as a predictive model, featuring the fact that it includes statistical noise coming from the $\delta$'s and $\eta$'s (see Figure~\ref{fig:Stat errors} and Eq.~\eqref{likelihood model}), not just the posterior uncertainty in the parameters, and we simply use Bayesian posterior prediction, namely

\begin{equation}\label{ppd}
    P(y_i^\text{pred} | \boldsymbol{Y}) = \int P(y_i^\text{pred} | \theta) P(\theta | \boldsymbol{Y}) \, d \theta,
\end{equation}
where $P(\theta | \boldsymbol{Y})$ is the posterior density of all model parameters. In this fashion, the posterior uncertainty on the parameters, and the statistical uncertainty from the likelihood model, are both taken into account in a principled way. Note that this predictive calculation can be performed for all 20 observables of interest, though ECP-type comparisons with the experimental datapoints happen only for the 18 points we have, excluding charge radii for $^{100} \text{Sn}$ and $^{68} \text{Ni}$. The next subsection explains how all Bayesian analyses are implemented numerically.

\subsection{Metropolis-Hastings and \texttt{surmise}}

The difficulty with any Bayesian method is to know how to understand the statistical properties of the posteriors. The simplest way to answer this question is to sample repeatedly (and efficiently) from those probability distributions. To sample from the posterior densities of the model parameters $\theta$, we use the standard Metropolis-Hastings algorithm implemented in the \texttt{surmise} Python package~\cite{surmise2021}.
For the results presented here, 8 independent chains of 725,000 samples were ran using the direct Bayes calibrator within \texttt{surmise}. For the step size used, the ratio of acceptance for proposed new steps was about $30\%$ across all chains. The first 100,000 samples of each chain were taken as a burn-in, reducing the effective sample size of the calibration to 5,000,000 evaluations.
The 8 chains were run in parallel and thus the 5,800,000 total evaluations took about a day on a personal computer with commodity hardware.
For comparison, it would have taken nearly six years of continuous computation to produce the same results using the original code for the high-fidelity model.

To evaluate posterior predictive distributions to compare with the experimental data, we rely on the fact that \texttt{surmise}, like any flexible Monte-Carlo Bayesian implementation, gives us access to all Metropolis-Hastings samples. For each multivariate sample of the parameters in $\theta$, we draw independent samples from the normal distributions in the likelihood model equation \eqref{likelihood model}, and evaluate the corresponding value of $y_i^\text{ex}$ in that model by plugging those sampled values into the right-hand side of that specification. Note that the second line in \eqref{likelihood model} must be used for this purpose, not the first line, in order to account for the uncertainty due to emulation. This procedure provides a sampling method for the distribution in \eqref{ppd} which has a level of accuracy consistent with the accuracy of the Metropolis-Hastings method for sampling from the parameters' posterior densities.

\section{Results and Discussion}\label{Sec: Results}

Having defined the covariant density functional model in Sec.~\ref{Sec:RMF}, the reduced basis emulator in Sec.~\ref{Sec: RBM}, and the statistical framework and the computational sampling tools in Sec.~\ref{Sec: Bayesian} we are in position to use the experimental data to calibrate the model under a Bayesian approach. At the time of the original calibration of
the FSUGold2 functional~\cite{Chen:2014sca}, this would have represented an exceptional computational challenge, mainly because of the absence of the computational speed up of three orders of magnitude provided by the RBM. Instead, in the original 
calibration, one was limited to finding the minimum of the objective (or $\chi^{2}$) function and the matrix of 
second derivatives. In this manner, it was possible to compute uncertainties and correlations between 
observables, but only in the Gaussian approximation. We compare and contrast our results and procedure, highlighting that with the exception of the information on the four giant monopole resonances and the maximum neutron star mass, both calibrations share the same dataset of binding energies and charge radii.

\begin{figure*}[]
    \centering
    \includegraphics[width=0.95\textwidth]{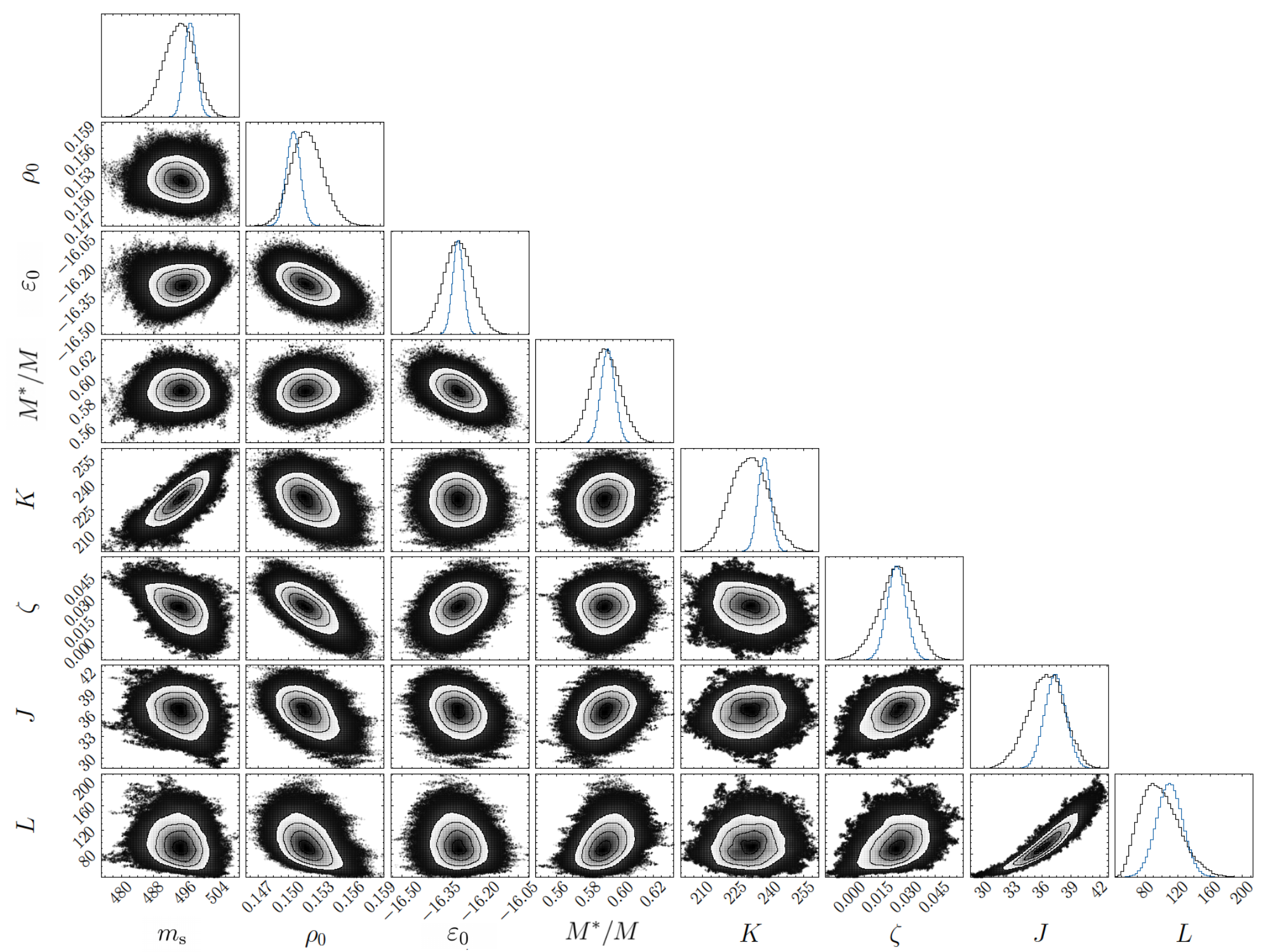}
    \caption{Corner plot \cite{corner} from the posterior distribution of the 8 bulk matter parameters $\theta$ obtained from the Metropolis-Hasting sampling with \texttt{surmise}. A total of 5 million samples were used, distributed along 8 independent chains. The saturation density $\rho_0$ is expressed in fm$^{-3}$, the mass of the $\sigma$ meson $m_\text{s}$, the binding energy at saturation $\varepsilon_0$, the  incompressibility coefficient $K$, the symmetry energy $J$ and its slope $L$ at saturation are all expressed in MeV. For comparison, the original calibration done in \cite{Chen:2014sca} is shown as the blue curves along the diagonal, scaled vertically to fit in the same plot as our posterior results in black.}
        \label{fig: CornerPlot}
\end{figure*}

We begin by displaying in graphical form the results of our Bayesian implementation in \texttt{surmise} as a corner plot in Figure~\ref{fig: CornerPlot}. The corner 
plot summarizes the posterior distribution of bulk parameters alongside the two-dimensional correlations.
For comparison, the Gaussian distribution of parameters extracted from the original FSUGold2 calibration is displayed by the vertically scaled blue line. As expected, the width of the one-dimensional distributions has
increased---in some cases significantly---relative to the Gaussian approximation that is limited to explore the
parameter landscape in the vicinity of the $\chi^{2}$ minimum. The inclusion of bigger estimated model errors $\delta_i$ in Eq.~\eqref{likelihood model} likely also share responsibility for the increased overall uncertainty.
Besides the increase in the width of the 
distribution, we see a relatively significant shift in the average value of the incompressibility coefficient $K$. 
We attribute this fact to the lack of information on the GMR, which is the observable that is mostly sensitive
to $K$.

\begin{figure*}[t]
    \centering
    \includegraphics[width=0.98\textwidth]{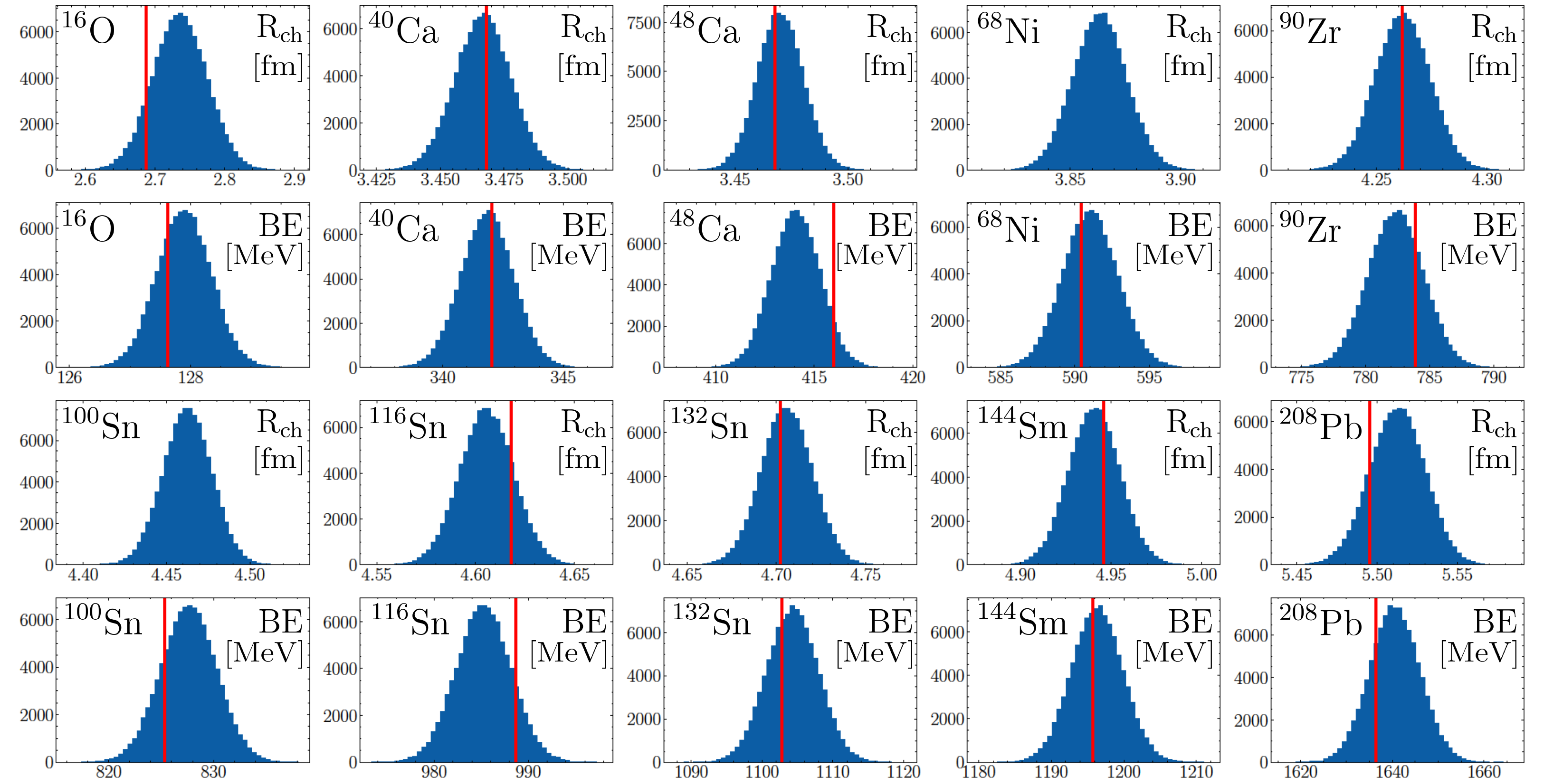}
    \caption{Posterior distributions for the binding energies (in MeV) and charge radii (in fm) of the ten nuclei involved in our study. A total of 100,000 samples from the 5,000,000 visited parameter values were used to make these distributions. The vertical red line in each plot represents the associated experimental values, contained within the model $95\%$ credible interval in all cases. Both $^{68}$Ni and $^{100}$Sn do not have a measured charge radius, making their associated posterior distributions true predictions from our calibration. The numerical values for the mean and credible intervals on all these quantities are displayed in Table~\ref{Tab: Credible Intervals}}
        \label{fig: Observables}
\end{figure*}

\begin{table*}[t]
\begin{tabular}{ccccccc}
Nucleus       & $\langle \text{R}^\text{em}_\text{ch}\rangle$ & $ [2.5\% -  97.5\%]$   & $\text{R}^\text{ex}_\text{ch}$ \  &  \  $\langle \text{BE}^\text{em}\rangle$ & $[2.5\% -  97.5\%]$    & $\text{BE}^\text{exp}$  \\ \hline
$^{16}$O   & 2.736      & [2.660  - 2.812] & 2.690  &\ \ 127.90       & [127.04 - 128.77] & 127.62  \\
$^{40}$Ca  & 3.467      & [3.446 - 3.488] & 3.471  &\ \ 341.83      & [339.75 - 343.91] & 342.05  \\
$^{48}$Ca  & 3.470       & [3.451 - 3.490]  & 3.470  &\ \ 414.05      & [411.66 - 416.45] & 416.00  \\
$^{68}$Ni  & 3.864      & [3.841 - 3.888] & -      &\ \ 590.99      & [587.47 - 594.52] & 590.41  \\
$^{90}$Zr  & 4.262      & [4.238 - 4.286] & 4.264  &\ \ 782.34      & [778.14 - 786.52] & 783.90  \\
$^{100}$Sn & 4.462      & [4.433 - 4.490] & -      &\ \ 827.69      & [822.49 - 832.87] & 825.30  \\
$^{116}$Sn & 4.606      & [4.580 - 4.632] & 4.620  &\ \ 985.21      & [979.57 - 990.86] & 988.68  \\
$^{132}$Sn & 4.705      & [4.678 - 4.733] & 4.704  &\ \ 1104.3      & [1097.3 - 1111.4] & 1102.84 \\
$^{144}$Sm & 4.941      &[4.914 - 4.968] & 4.947  &\ \ 1196.3      & [1189.4 - 1203.1] & 1195.73 \\
$^{208}$Pb & 5.512      &[5.478 - 5.544] & 5.497  &\ \ 1640.7      & [1630.1 - 1651.3] & 1636.43
\end{tabular}\caption{Mean values and $95\%$ credible intervals of the Bayesian posteriors on charge radii (in fm) and binding energy (in MeV), showed in Figure~\ref{fig: Observables}. Also displayed are the 18 available experimental values \cite{Chen:2014sca}. The credible intervals are calculated as equal-tailed intervals---such that the probabilities of falling above or below the interval are both equal to 2.5\%.  }\label{Tab: Credible Intervals}

\end{table*}

Beyond the corner plot that displays the distribution of bulk parameters and the correlations among 
them, we illustrate in Fig.\ref{fig: Observables} and Table~\ref{Tab: Credible Intervals} the performance of the model as compared with the 
experimental data informing the calibration. The blue histograms display the posterior predictive 
distributions~\eqref{ppd} of each of the 20 observables. Included in our results are predictions for the yet to be
measured charge radii of $^{68}$Ni and $^{100}$Sn. The vertical red lines indicate the values of the 
experimental datapoints specified in\,\cite{Chen:2014sca}. These plots show excellent coverage of 
all datapoints within our reported uncertainty. With 18 datapoints, one would expect about one of
them to fall outside of 95\% credible intervals. The credible intervals are printed in Table~\ref{Tab: Credible Intervals}, showing that none of our datapoints fall outside those intervals around the posterior means.
This implies that our uncertainty quantification leans towards the conservative side, although the binding energy for $^{48}$Ca is very close to falling outside the 95\% band. 
This is reassuring, as a strong indication that our uncertainty quantification is only moderately conservative. 
Our method has produced uncertainties which are very likely not to be overly wasteful by significantly 
over-reporting uncertainty, and which are very likely not to under-report uncertainty. This is exactly where 
a Bayesian predictive posterior coverage analysis wants to be, in a study with such a small number of 
datapoints. 

Being the lightest of all the nuclei included in the calibration, $^{16}$O may be regarded as a questionable 
``mean-field" nucleus. As such, comparing its experimental charge radius with our posterior results is particularly interesting, since the model standard deviation 
we used was more than 5 times larger than for the other observables. Yet, our reported uncertainty sees the 
experimental measurement fairly well reproduced. This is an indication that it was important to use the 
higher model variance, or our prediction could have reported too low of an uncertainty. Using a smaller variance for the model error $\delta_i$ could have also pushed the parameters too strongly towards the $^{16}$O charge radius outlier, deteriorating 
the overall performance of the calibration on the other nuclei. The final coverage of all data points illustrates our method's ability 
to handle heteroskedasticity (uneven variances) well. Finally in Figure~\ref{fig: Observables}, because our comparison with predictive 
distributions performs very well and is only mildly conservative, we can be confident that our predictions 
for the charge radii of both $^{68}$Ni and $^{100}$Sn are robust. How narrow these histograms are is a 
testament to the quality of the original modeling, its emulation, and our uncertainty quantification.

Coming back to the corner plot in Figure~\ref{fig: CornerPlot}, we note that the strongest correlation between observables involves the value of the 
symmetry energy ($J$) and its slope ($L$) at saturation density. The symmetry energy quantifies the energy cost in transforming symmetric nuclear matter---with equal 
number of neutrons and protons---to pure neutron matter. In the vicinity of nuclear matter saturation density, 
one can expand the symmetry energy in terms of a few bulk parameters\,\cite{Piekarewicz:2008nh}: 
\begin{equation}
 {\cal S}(\rho) = J + Lx + \frac{1}{2}K_{\rm sym}x^{2}+\ldots 
 \label{SymmE}
\end{equation}
where $x\!=\!(\rho\!-\!\rho_{0})\!/3\rho_{0}$ is a dimensionless parameter that quantifies the deviations 
of the density from its value at saturation. Given that the calibration is informed by the binding energy
of neutron-rich nuclei, such as ${}^{132}$Sn and ${}^{208}$Pb, the symmetry energy 
$\widetilde{J}\!=\!S(\widetilde{\rho})\!\approx\!26\,{\rm MeV}$  is well constrained at an average density 
of about two thirds of saturation density, or $\widetilde{\rho}\!\approx\!0.1\,{\rm fm}^{-3}$\,\cite{Furnstahl:2001un}.
As a result, one obtains the following relation:
\begin{equation}
 \widetilde{J} = J - \frac{L}{9} + \frac{K_{\rm sym}}{162} + \ldots
 \approx  J - \frac{L}{9} \implies J \approx \widetilde{J} + \frac{L}{9} \;.
\label{JL}
\end{equation}
Hence, accurately calibrated EDFs display a strong correlation between $J$ and $L$ by the mere fact
that the calibration included information on the binding energy of neutron-rich nuclei. In the original
calibration of FSUGold2\,\cite{Chen:2014sca}, one obtained a correlation coefficient between $J$ and 
$L$ of 0.97, while in this work we obtained a correlation coefficient of 0.92. The slight non-linearity observed in Fig.\ref{fig: CornerPlot} on the correlation between $J$ and $L$ is due to $K_{\rm sym}$, which was neglected in the 
simple argument made in Eq.(\ref{JL}).

\begin{figure}[t]
    \centering
    \includegraphics[width=0.45\textwidth]{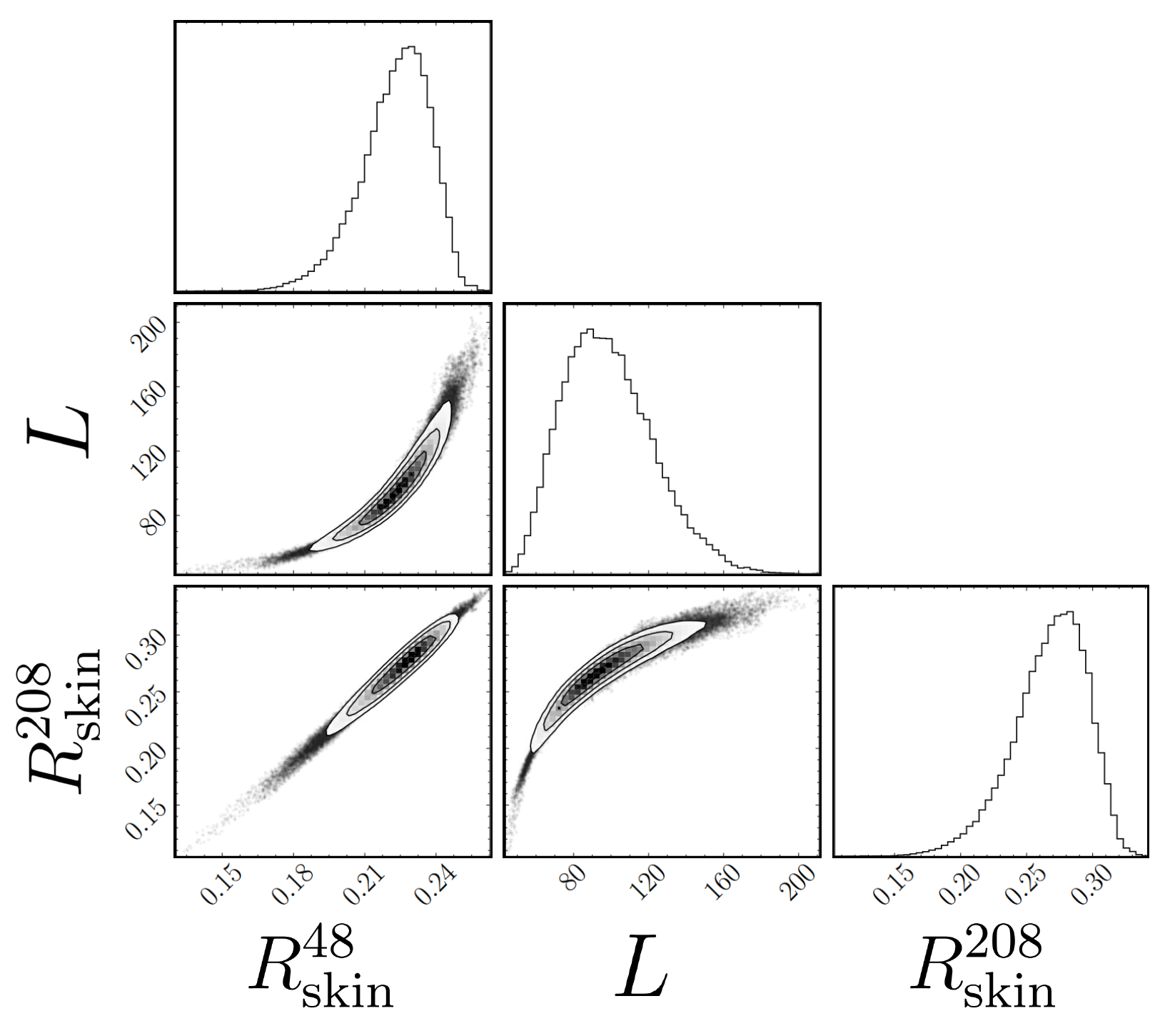}
    \caption{Correlation corner plot \cite{corner} between the posterior distributions for the neutron skin of $^{48}$Ca and $^{208}$Pb (both in fm), and the slope of the symmetry energy $L$ (in MeV). A total of 100,000 samples from the 5,000,000 visited parameter values were used to make these distributions. Both neutron skins $R_{\rm skin}^{48}$ and $R_{\rm skin}^{208}$ are strongly correlated, each with negative skewness. The distribution for $L$, on the other hand, has a positive skewness and, while it is strongly correlated with both neutron skins, the correlation displays a non-linear behavior in both cases.}
        \label{fig: L}
\end{figure}

The slope of the symmetry energy may also be extracted from laboratory experiments. Although not 
directly an observable, $L$ has been determined to be strongly correlated to the thickness of the neutron 
skin of heavy nuclei\,\cite{Brown:2000,Furnstahl:2001un,RocaMaza:2011pm}; the neutron skin thickness
is defined as the difference in the mean square radii between the neutron and proton vector densities (see Eqs.~\eqref{Eq: Radius}). In Fig~\ref{fig: L} we show the correlation plot between $L$ and the neutron skin of $^{48}$Ca and $^{208}$Pb calculated directly from 100,000 random samples from our posterior distributions. It is important to note that we have not included the model error $\delta_i$ through Eq.~\eqref{ppd} in these histograms \footnote{Such procedure would require to first give an accurate estimation of the model error on \emph{neutron} radii -a non trivial task given the lack of experimental data-, and second to take into account the model correlation between $R_p$ and $R_n$.}, and as such we don't expect the uncertainties to be accurate, as we discuss later in Sec.~\ref{Sec: Conclusions and outlook}.

The correlation between $L$ and the thickness of the neutron skin of heavy nuclei has a strong physical underpinning. For example, in the case of ${}^{208}$Pb, surface tension 
favors the formation of a spherical liquid drop containing all 208 nucleons. However, the symmetry energy increases monotonically in the density region of relevance to atomic nuclei. Hence, to minimize the symmetry
energy, is energetically favorable to move some of the excess neutrons to the surface. It is then the difference
in the symmetry energy at the core relative to its value at the surface that determines the thickness of the
neutron skin; such a difference is encoded in the slope of the symmetry energy $L$. If such a difference is
large enough to overcome surface tension, then some of the excess neutrons will be pushed to the surface,
resulting in a thick neutron skin\,\cite{Horowitz:2001ya}. 
That the correlation between the neutron skin thickness of ${}^{208}$Pb and $L$ is strong has been validated 
using a large set of EDFs\,\cite{RocaMaza:2011pm}. Note that $L$ is closely related to the pressure of pure neutron 
matter at saturation density---a quantity that has been extensively studied using chiral effective field 
theory\,\cite{Hebeler:2009iv,Tews:2012fj,Kruger:2013kua,Lonardoni:2019ypg,Drischler:2021kxf,
Sammarruca:2021mhv,Sammarruca:2022ser}, and which is of great relevance to our understanding
of the structure of neutron stars\,\cite{Piekarewicz:2019ahf}.

The Lead Radius Experiment (PREX) reported the following value for the neutron skin thickness of ${}^{208}$Pb\,\cite{Adhikari:2021phr}:
\begin{equation}
 R_{\rm skin}^{208}=(0.283\pm0.071)\,{\rm fm},
 \label{Rskin208}
\end {equation} 
Although there is some mild model dependence in the extraction of $R_{\rm skin}^{208}$ from the model 
independent electroweak measurement of the parity violating asymmetry, such a drawback will have no
impact on our discussion of Fig.\,\ref{fig: L}. However, a more faithful analysis should assess the impact 
of the true physical observable---the parity violating asymmetry or the weak-charge form factor---on the density
dependence of the symmetry energy\,\cite{Reinhard:2022inh}.
For now, we will compare with Ref. \cite{Reed:2021nqk} that inferred the large value of $L\!=\!(106 \pm 37)\,{\rm MeV}$\, by relying on the strong correlation 
between the neutron skin thickness of ${}^{208}$Pb and the slope of the symmetry energy alluded to earlier\,\cite{RocaMaza:2011pm}.
Such a large value of $L$ systematically overestimates existing 
limits based on both theoretical approaches and experimental measurements involving hadronic 
probes\,\cite{Reed:2021nqk}. By examining the distributions for both $L$ and  
$R_{\rm skin}^{208}$, we observe that they both fall comfortably within the experimental range. Indeed, for the 100,000 
samples from our parameters' posterior distribution used in the figure, the mean value for $R_{\rm skin}^{208}$ is $0.267$ fm with a standard deviation of $0.028$ fm, while the mean value for $L$ is $98.70$ MeV with a standard deviation of $23.77$ MeV. It is 
important to note that no information on neutron skins---or any other observable that is strongly
correlated to $L$---was included in the calibration procedure, making it difficult to estimate the model error associated with such quantities. This also indicates that in the absence 
of any guidance, the class of covariant EDFs used in this work tend to produce stiff symmetry energies, in contrast to chiral effective field theories that tend to favor relatively soft symmetry 
energies\,\cite{Hebeler:2009iv,Tews:2012fj,Kruger:2013kua,Lonardoni:2019ypg,Drischler:2021kxf,
Sammarruca:2021mhv,Sammarruca:2022ser}. Particularly interesting to note is that whereas 
$R_{\rm skin}^{208}$ and $L$ are strongly correlated, the correlation deviates significantly from
the one obtained using a large set of both covariant and Skyrme energy density 
functionals\,\cite{RocaMaza:2011pm}. It is known, however, that $R_{\rm skin}^{208}$ displays a stronger correlation with the slope of the symmetry energy at $0.1\,{\rm fm}^{-3}$ than at $\rho_{0}$; see Ref.\cite{Reed:2021nqk} and references contained therein.

However, the correlation between $R_{\rm skin}^{208}$ and $R_{\rm skin}^{48}$ remains as strong 
as observed in Ref.\,\cite{Piekarewicz:2021jte}. Such a strong correlation suggests that, if correct,
the posterior distribution for the neutron skin thickness of $^{48}$Ca should peak around 
$R_{\rm skin}^{48}\!\approx\!0.23\,{\rm fm}$. This assertion is consistent both with Fig.\,\ref{fig: L} as
well as with Ref.\,\cite{Chen:2014sca}. For the 100,000 samples from our parameters' posterior distribution used in the figure, the mean value for $R_{\rm skin}^{48}$ is $0.223$ fm with a standard deviation of $0.015$ fm. However, this seemingly straightforward deduction is in stark 
disagreement with experiment. Indeed, the CREX collaboration has recently reported a neutron skin 
thickness that is significantly smaller than the theoretical prediction\,\cite{CREX:2022kgg}:
\begin{equation}
 R_{\rm skin}^{48}=(0.121\pm0.035)\,{\rm fm}.
 \label{Rskin48}
\end {equation} 

It becomes abundantly clear that the physics encapsulated in the Lagrangian density depicted
in Eq.(\ref{LDensity}) is insufficient to describe simultaneously the neutron skin thickness of both
$^{48}$Ca and $^{208}$Pb. Granted, with only two isovector parameters the model may be too
rigid to break the strong observed model correlation between $R_{\rm skin}^{208}$ and $R_{\rm skin}^{48}$. 
However, whereas models with a more refined isovector sector may be able to reconcile both 
measurements at some level, a consensus is emerging that this can only be done at the expense 
of introducing significant tension
with other calculated observables by the model~\cite{Hu:2021trw,Reinhard:2022inh,Zhang:2022bni,
Mondal:2022cva}. From the limited perspective of covariant DFT, this tension is not surprising. 
As long as the neutron skin develops from the competition between surface tension and the slope 
of the symmetry energy, a correlation between the neutron skin thickness of neutron-rich nuclei 
is unavoidable. 

The existing tensions from parity violating electron scattering experimental results, as well as the anticipated new laboratory experiments and astronomical observations that will be coming in the next years, will demand more efficient and sophisticated approaches to develop extensions to the model. As an immediate direction, we are contemplating the calibration of covariant EDFs with a more elaborated isovector sector that might help bridge both neutron skin results without compromising the success of the model in reproducing other nuclear observables, such as the ones displayed in Fig~\ref{fig: Observables}. As we discuss in the next and final section, well quantified uncertainties enabled by powerful emulators such as the RBM will be indispensable to achieve those goals.

\section{Conclusions and Outlook} \label{Sec: Conclusions and outlook}

In the last few decades nuclear theory has gone through several transformational changes brought on by embracing 
philosophies and techniques from the fields of statistics and computational science.
It is now expected that theoretical predictions should always be accompanied by theoretical
uncertainties\,\cite{join2011uncertainty}.
This is particularly true in theoretical nuclear physics where predictions
from QCD inspired models require the calibration of several model parameters.
This newly-adopted philosophy has also prompted the exploration of uncertainty quantification 
across the many sub-fields of nuclear theory\,\cite{godbey2022,king2019direct,odell2022performing,
mcdonnell2015uncertainty,drischler2020well}.
Furthermore, several recent advancements and discoveries have become feasible only through the successful 
integration of machine learning and other novel computational approaches to the large body of theoretical 
models developed over many decades\,\cite{boehnlein2022colloquium,hamaker2021precision,
utama2016nuclear,kuchera2019machine,neufcourt2019neutron}.
This dedication is also exemplified by the theory community's proactive efforts to organize topical conferences, 
summer schools, and workshops in service of disseminating the technical know-how to every level of the 
community.

Aligned with these developments and efforts, our present work aimed at showcasing a pipeline for integrating a statistical framework through one such innovative computational technique. We have calibrated a covariant energy density functional within a Bayesian approach using available experimental values of binding energies and charge radii. The calibration of the model, as well as the quantification of the uncertainties of its predictions, required millions of evaluations for different values of its parameters. Such titanic computational burden was made possible---straightforward even---thanks to the emulation of the model through the reduced basis method, which decreased the necessary calculation time from months or years to a single day on a personal computer.

Our calibration's main results, which consists of posterior distributions for all the model's parameters, were presented in Figure~\ref{fig: CornerPlot}. From these posteriors, and following the statistical framework we developed in Sec.\,\ref{Sec: Bayesian}, the model output can be estimated with well quantified uncertainties that can take into account experimental, model, and emulator errors. We showed such calculations with their respective estimated uncertainties in Figure~\ref{fig: Observables} and Table~\ref{Tab: Credible Intervals} for the binding energies and charge radii of the 10 nuclei involved in the study, as well as the model's posterior distributions for the neutron skin of $^{48}$Ca and $^{208}$Pb in Figure~\ref{fig: L}. The fact that the experimental values used in the calibration, depicted as red vertical lines in Figure~\ref{fig: Observables}, fall within the $95\%$ our calculated credible intervals gives us confidence that our uncertainty procedure was not biased towards being too optimistic for this dataset. This is especially true for the case of the charge radii of $^{16}$O, which was treated as an outlier based on prior expert knowledge on the expectation of the limits of the mean field approach for smaller systems. Once the experimental values for the charge radii of $^{68}$Ni and $^{100}$Sn become available, it will be interesting to contrast our model prediction's and gauge the success of the uncertainty level estimated.

However, the picture changes when we focus on the predictions for the neutron skin thickness of $^{48}$Ca and $^{208}$Pb showed in 
Fig.\,\ref{fig: L}. The recent experimental campaigns PREX\,\cite{abrahamyan2012measurement}, 
PREX-II\,\cite{adhikari2021accurate}, and CREX\,\cite{CREX:2022kgg} on parity violating electron scattering have published 
results which suggest that the neutron skins of $^{48}$Ca and $^{208}$Pb stand in opposite corners. While $^{208}$Pb is 
estimated to have a relatively thick neutron skin of around $0.28$ fm\,\cite{adhikari2021accurate}, $^{48}$Ca\,\cite{CREX:2022kgg} 
is estimated to have a significantly smaller skin of around $0.12$ fm. As our model currently stands, its posterior calculations 
are unable to satisfy both values simultaneously. 

We believe that the cause of this tension is twofold. First, the scarcity of experimental information on strong isovector indicators 
hinders the amount of information that can be incorporated into the definition of the likelihood given in Sec.\,\ref{subsec: Like}. 
Second, precisely because of the lack of experimental information, there was no need---until now---to extend the isovector 
sector of the model beyond two isovector parameters. As such, the present EDF may be too rigid to reproduce the neutron skin
thickness of both $^{48}$Ca and $^{208}$Pb without compromising the success of the model in reproducing other 
well-measured observables.

Moving forward, we envision two complementary research directions that could help mitigate the problems identified above.
First, one could build a more robust statistical framework that, by including strong isovector indicators, such as
information on the electric dipole response of neutron rich nuclei, will impose stringent constraints on the isovector 
sector. Second, and as already mentioned, we could increase the flexibility of the isovector sector by adding additional
interactions that modify the density dependence of the symmetry energy. The use of dimensionality reduction techniques 
such as the RBM to significantly speed up the calculation time---especially if information on nuclear excitations is 
incorporated into the calibration of the EDF---will become a fundamental pillar of the fitting protocol.

We believe that the RBM we showcased here has the potential to further impact many of the nuclear theory areas that have already made use of similar emulators, as well as expanding the frontiers 
of the physical models that can be successfully emulated. Indeed, the RBM's unique combination of few high-fidelity evaluations needed to build an effective emulator, the simplicity and flexibility of the Galerkin projection, and the ability to precompute many observables and equations in the offline stage could
allow the community to deploy trained emulators for use on different computer architectures and on cloud infrastructure~\cite{bmex}. This could effectively lower the barrier created by the need of running expensive computer models locally. This could give access of cutting edge theoretical models and simulations to an increased number of research groups, opening new opportunities to expand the network of collaborative research. 

In short, the computational framework detailed in this work attempts to provide an end-to-end solution for model calibration and exploration with a focus on statistical rigor without sacrificing computational efficiency.
By leveraging this efficiency to nimbly incorporate new experimental data, one can imagine the continuous calibration of models that can be updated in a matter of hours without requiring large-scale computing facilities.
Finally, the heavy focus on integrating these disparate parts into a user-friendly form to generate physics-informed emulators is ultimately in service of our wider goal to increase data availability and software accessibility, and is a necessity in the paradigmatic shift towards probability distributions---rooted in Bayesian principles---defining physical models rather than a single set of optimal parameters.

\section*{Conflict of Interest Statement}

The authors declare that the research was conducted in the absence of any commercial or financial relationships that could be construed as a potential conflict of interest.

\section*{Funding}
This work was supported by the National Science Foundation CSSI program under award number 2004601 (BAND collaboration) and the U.S. Department of Energy under Award Numbers DOE-DE-NA0004074 (NNSA, the Stewardship Science Academic Alliances program), DE-SC0013365 (Office of Science), and DE-SC0018083 (Office of Science, NUCLEI SciDAC-4 collaboration).
This material is based upon work supported by the U.S. Department of Energy Office 
 of Science, Office of Nuclear Physics under Award Number DE-FG02-92ER40750. 
 
\section*{Acknowledgements}
We are grateful to Moses Chan for guiding us on the use of the \texttt{surmise} python package, as well as for useful discussions and recommendations on good practices for the Monte Carlo sampling. We are grateful to Craig Gross for useful discussions about the reduced basis method, including the visualization of the speed up gained in terms of the high fidelity solver size $\mathcal{N}$. We thank Diogenes Figueroa for his read of the manuscript. 


\section*{Data Availability Statement}
The datasets generated and analyzed for this study can be found in the repository located \href{https://figshare.com/projects/RBM_Calibration_Pipeline/149840}{here}. The software framework that generated the RBM emulator, performs the calibration, and produces the raw plots can be found \href{https://github.com/kylegodbey/RMF-RBM}{here}.

\bibliographystyle{apsrev4-1}
\bibliography{ReferencesJP,Bibliography}

\end{document}